\documentclass[twocolumn,superscriptaddress,floatfix,pre,reprint]{revtex4-1}
\usepackage{graphicx}  
\usepackage{color}
\usepackage{dcolumn}   
\usepackage{bm}        
\usepackage{amssymb}   
\usepackage{amsmath}
\usepackage{color}
\usepackage{float}
\usepackage{upgreek}

\usepackage{epsfig}
\usepackage{latexsym}
\usepackage{natbib}
\usepackage{subfigure}
\usepackage{textcomp}

\newcommand{\HIU}{Helmholtz Institute for Electrochemical Energy Storage Ulm, D--89073 Ulm, Germany}
\newcommand{\ULM}{Institute of Theoretical Chemistry, Ulm University, D--89073 Ulm, Germany}

\begin{document}

\title{Atomistic modeling of Li- and post-Li ion batteries}
\author{H.\, Euchner}\affiliation{\HIU}
\author{A.\, Groß}\affiliation{\HIU}\affiliation{\ULM}


\begin{abstract}
Alkali metal ion batteries, and in particular Li--ion batteries, have become a key technology for current and future energy storage, already nowadays powering many devices of our daily lives. Due to the inherent complexity of batteries and their components, the use of computational approaches on all length and time scales has been largely evolving within recent years. Gaining insight in complex processes or predicting new materials for specific applications are two of the main perspectives computational studies can offer, making them a indispensable tool of modern material science and hence battery research. After a short introduction to battery technology, this review will first focus on the theoretical concepts that underlie the functioning of Li-- and post--Li--ion batteries. This will be followed by a discussion of the most prominent computational methods and their applications, currently available for the investigation of battery materials on an atomistic scale.  
\end{abstract}

\maketitle

\section{Introduction}  


Already today our society is facing enormous challenges with respect to global warming and its consequences. Stopping or at least slowing down climate change, however, means that we have to change the way we are producing and consuming energy.  While renewable energies are an indispensable part to achieve a zero emission society, their intermittent character calls for advanced energy storage concepts. Therefore, apart from their use in electronic devices and for e--mobility, batteries have become one of the key technologies for our near future~\cite{Dunn2011,Goodenough2014}. 
Currently the battery market is dominated by Li-ion technology, however, for applications such as stationary storage post Li--ion technologies are recently gaining significantly more interest, as \textit{e.g.} sodium is much cheaper and more easily available than lithium. This is likely to become an important factor with respect to the increasing demand for energy storage in our society~\cite{Yabuuchi2014,Chayambuka2018,Vaalma2018}.
Hence, it is no exaggeration to emphasize the importance of battery technology and the further development of the different aspects of a functioning battery. Indeed, since the commercialization of the first Li--ion batteries (LIBs) in 1991~\cite{Nishi2001}, there has been a tremendous improvement in battery performance, which, on the other hand, results in steadily increasing requirements that batteries have to fulfill~\cite{Palacin2009,Thackeray2012,Goodenough2014,Nitta2015,Li2018, Ma2020}. Different kinds of applications call for higher volumetric and gravimetric energy density, high rate capability, long cycle life, safety and of course low cost~\cite{Goodenough2010}. 
In the quest for batteries with a better performance, theoretical studies addressing structures and processes in batteries on an atomic level play an increasingly important role~\cite{Lee2014, Hoermann2015, Saubanere2016, Gross2018TCC, Yahia2019, Liu2020, Ma2021}. In this review, we will describe the theoretical and numerical methods employed in the evaluation of battery properties and also highlight the insights gained from such studies.
This review starts with a short discussion of state of the art battery technology and introduces some of the standard battery terminology, typically using LIBs as prototypical example. In the next chapters, the underlying theoretical concepts of a battery will be derived, before the most prominent computational approaches that are currently used in battery research will be discussed. While we are often referring to specific materials when discussing particular numerical methods, this review has its focus on theoretical concepts and computational approaches, 
thus not meaning to give a comprehensive overview over recent developments in the battery field. Note also that many aspects of the approaches that are presented here can be generalized to all types of batteries, the focus of this work, however, lies on Li-- and post--Li--ion systems and their specific challenges. 

\subsection{State of the art}

\begin{figure}[b]
 \center\includegraphics[angle=0,width=0.8\columnwidth]{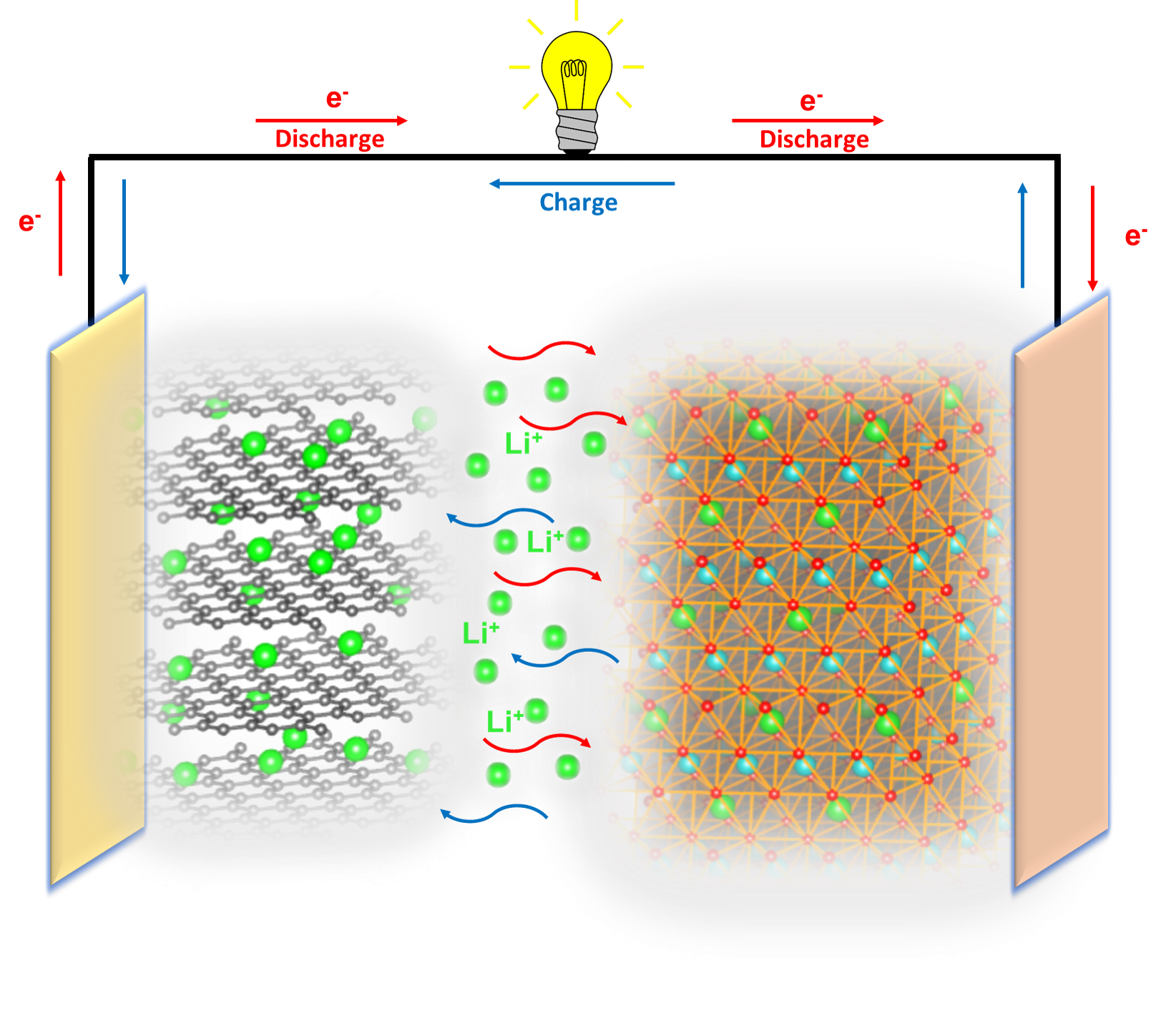}
 \caption{Schematic of the functioning of a rocking chair type Li--ion battery with a graphite anode and a TM--oxide cathode. During discharge, Li--ions are de--intercalated at the anode and intercalated at the cathode, while the electrolyte is responsible for shutteling the ions between the electrodes.}
 \label{fig:battery}
\end{figure}

In a prototypical Li--ion cell that is nowadays in use, a likely setup consists of a graphitic anode and a NMC (Ni--Mn--Co oxide) type cathode, which are separated by a membrane soaked with electrolyte that shuttles the Li--ions between anode and cathode (see Fig.~\ref{fig:battery}).
During discharge of such a rocking chair type LIB positively charged ions are deintercalated at the anode and shuttle to the cathode, where they intercalate inbetween the layers of the NMC. The corresponding electrons take the path through the external electric circuit, thus allowing for the exploitation of the energy gain of the underlying reaction -- often also expressed as difference in the Li chemical potential between anode and cathode as will be demonstraded below -- for \textit{e.g.}, powering a electronic device. 

For battery applications, the amount of energy that can be stored is the crucial variable, as it for instance determines how far an electric vehicle can drive. The ability to store energy is quantified by the energy density which is either given as energy per unit weight (gravimetric or specific energy density, in Wh/kg) or per unit volume (volumetric energy density, in Wh/l). 
With respect to improving the energy density, the cathode is the most decisive factor, resulting in large research efforts aiming at improved and new cathode materials. The energy density of a cathode is determined by its storage capacity and its operating voltage. While the capacity determines the number of ions (and electrons) that can be stored in an electrode per unit weight (gravimetric or specific capacity in mAh/g) or per unit volume (volumetric capacity in mAh/l), the operating voltage states the potential of the cathode with respect to a chosen reference (typically the respective alkali metal).

\begin{figure}
 \center\includegraphics[angle=0,width=0.8\columnwidth]{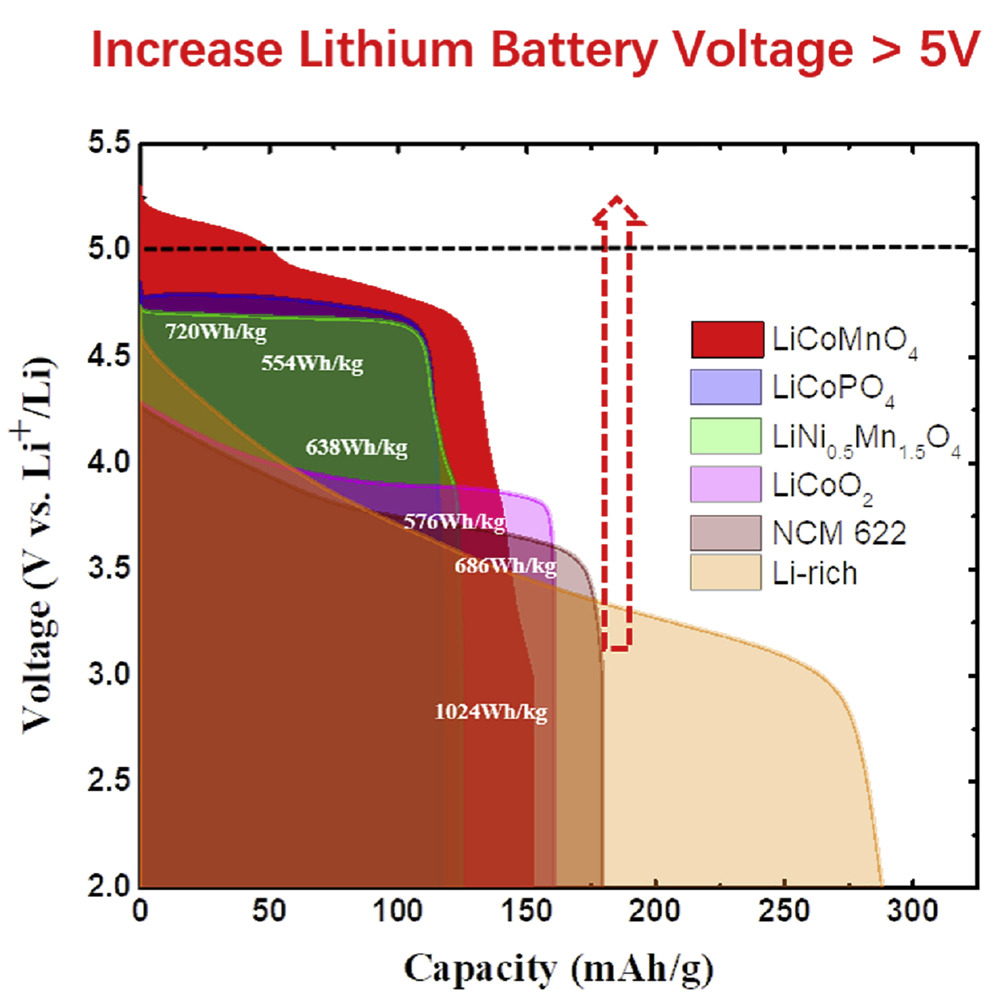}
 \caption{Typical discharge curves for state of the art cathode materials in LIBs. The areas below the curves yield the corresponding energy densities. From Chen et al.~\cite{Chen2019a}. With permission from Elsevier.}
 \label{fig:Chen}
\end{figure}

In Fig.~\ref{fig:Chen} the voltage profile of typical cathode materials is depicted as a function of the capacity. The energy density of these cathode materials is obtained from the integral in Eq.~(\ref{fig:Chen}) and corresponds to the area below the voltage profile:
\begin{equation}
 E = \int U(q) dq,
\end{equation}
with $U(q)$ the voltage as a function of the capacity. 
Clearly, significant progress has been achieved during the last decades. The originally introduced layered LiCoO$_2$ electrodes have further evolved and cobalt has partially been replaced by more environmentally benign materials such as manganese and nickel. This resulted in the development of
the so–-called NMC cathodes (Li$_x$Ni$_y$Mn$_z$Co$_{1-y-z}$O$_2$)~\cite{Rozier2015}, which are typically named with respect to the atomic ratio of the transition metals (\textit{e.g.}, NMC532, NMC622,
NMC811, etc.). A next step for achieving higher energy densities was to aim at an increase of the Li–-content in these layered cathode materials, which resulted in the development of Li-–rich NMCs~\cite{Rozier2015}.
These achieve capacities beyond 250 mAh/g and reach energy densities of up to $\approx$1000 Wh/kg.
Moreover, other compounds such as spinel phases and polyanionic compounds have been developed.  

\begin{figure}
 \center\includegraphics[angle=0,width=0.75\columnwidth]{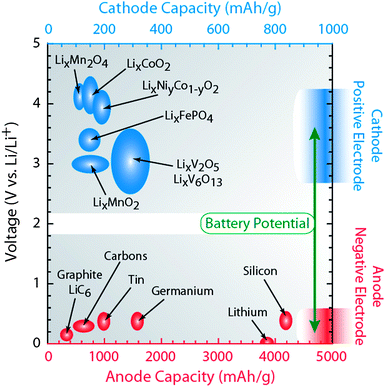}
 \caption{Cell voltage and capacity with respect to different anode/cathode combinations for LIBs. From Landi et al.~\cite{Landi2009}. With permission of the Royal Society of Chemistry.}
 \label{fig:Landi}
\end{figure}

In fact, while most efforts aim on improving cathode materials, the energy density nevertheless is determined by capacity and voltage (potential) of both electrodes, as can be inferred from Fig.~\ref{fig:Landi}. In general, the ideal battery anode lies at low potential and offers high capacity, whereas the ideal cathode offers a high voltage also combined with a high capacity. Indeed, in the case of LIBs Li--metal would be the ideal anode, however, safety issues related to the growth of dendrites have so far hindered the use of Li--metal as anode material in rechargeable LIBs~\cite{Wang2020}. 

\begin{figure}
 \center\includegraphics[angle=0,width=0.8\columnwidth]{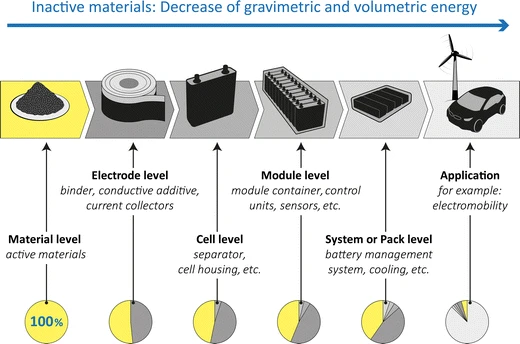}
 \caption{Interrelation of the energy density on material level to that on pack level and final application. From Placke et al.~\cite{Placke2017}. Reprinted with permission from this reference with permission from Springer Nature.}
 \label{fig:Placke}
\end{figure}

When discussing energy densities, one has to be careful which numbers are actually to be compared. The above discussed cathode energy densities are obtained on a material level, \textit{i.e.} only the active material is considered.  
However, batteries are complex many--component systems, meaning that further parts such as current collectors, binders and other additives are also involved, resulting in additional weight, going along with a highly complex interplay of these components~\cite{Wachtler2001,Qi2013,Kramer2013,Placke2017}. Consequently, the energy density of a working electrode is also influenced by these components, thus reducing its actual energy density, as schematically depicted in Fig.~\ref{fig:Placke}. 
Finally, when the performance of a full battery is assessed, it still makes a difference if one is referring to cell level, module level or even pack level (see Fig.~\ref{fig:Placke}).
Hence, while the above discussed active materials reach energy densities in the order of 1000 Wh/kg, the best commercially available Li--ion batteries currently reach about 260 Wh/kg on cell level \cite{Lobberding2020}.

The complexity of a battery -- \textit{i.e.} the different components and their interplay -- and the almost infinite number of possibilities for combining these components point to the importance of knowledge--based design strategies for next generation battery systems. Consequently, a first crucial step is the detailed investigation of the main components, where all attempts for understanding the whole battery system with the underlying mechanisms have to set in. Therefore, the typical approach, also on the experimental side, is the separate search for novel and improved anode, cathode and electrolyte materials. Yet, it has to be kept in mind that it is not only the single components but their interplay which makes a battery work, such that full cell studies are always needed. From a theoretical point of view, the simulation of a full battery on an atomistic scale is anyway still far from what currently can be achieved, such that the properties of anode, cathode and electrolyte have to be studied independently, or simplified model systems have to be addressed, where, \textit{e.g.}, only the interaction of electrolyte molecules with the electrode surface is investigated~\cite{Buchner2018,Buchner2019}.

\section{Electrochemical energy storage}
When Li--ion and post--Li-ion batteries -- with alkali metal ions shutteling back and forth between anode and cathode -- are discussed, one of the big advantages that facilitates the computational treatment is the fact that the ion leaving on the anode side is of the same type as the ion entering at the cathode side. This represents an important difference to classical battery types such as, \textit{e.g.}, the Daniell element. In fact, during the discharge of a Daniell element, at the anode Zn$^{2+}$ ions go in solution, while at the cathode side Cu is deposited. Hence, the differences in the respective solvation energies also contribute to the overall reaction and, therefore, to the voltage of the battery. 

\subsection{The general case}
While the fact that solvation free energies do not need to be accounted for make the computational treatment of standard Li-- and post--Li--ion systems much easier, we nevertheless first derive the battery voltage from a formal thermodynamic approach. By expressing the change in Gibbs free energy in terms of the thermodynamic variables, one obtains:
\begin{equation}
 dG \ = \ -SdT \ + \ Vdp \ + \ \sum_i \mu_i dn_i, 
\end{equation}

\noindent with $S$ the entropy, $T$ the temperature and $V$ and $p$ representing volume and pressure. Finally,
$\mu_i$ denotes the chemical potential of component $i$, while $dn_i$ refers to the corresponding change in the number of particles.
Yet, in an electrochemical environment, we are dealing with charged species and therefore, here the electrochemical potential $\tilde{\mu}_i = \frac{\partial G}{\partial n_i}$ has to be considered. The latter one can simply be expressed as the sum of chemical potential $\mu_i$ and electrostatic potential $\upvarphi_i$ multiplied by the corresponding charge $z_i e$:

\begin{equation}
\tilde{\mu}_i \ = \ \mu_i \ + \ z_i e \upvarphi_i \ = \ \mu_i^0 \ + \ k_BTln(a_i) \ + \ z_i e \upvarphi_i,
\label{eq:chem}
\end{equation}

\noindent where in the last step, the chemical potential of an ideal solution is introduced, which comprises the activity coefficients $a_i$ and the standard chemical potential $\mu_i^0$.
Now, we first consider a half cell (consisting of electrode and electrolyte) for which equilibrium conditions require that $dG = 0$. 
Hence, under the assumption of constant temperature and constant pressure, the following equation needs to be fulfilled for the half cell in equilibrium:

\begin{equation}
dG = \sum_i\tilde{\mu}_i dn_i = 0
\label{eq:half}
\end{equation}

In this formulation, $dn_i$ accounts for the change in particle number of the respective species involved. Considering a typical half cell reaction, such as $M \rightleftharpoons M^{z+} + ze^{-}$, this then translates into:

\begin{align}
dG &= \mathrel{\phantom{+}}\mu_M^0 \ + \ k_BTln(a_M) \nonumber \\
      &\mathrel{\phantom{=}} -\mu_{M^{z+}}^0 \ - \ k_BTln(a_{M^{z+}}) \ - \ z e \upvarphi_{sol} \nonumber\\
      &\mathrel{\phantom{=}} -n_e\mu_{e^-}^0 \ -  \ n_ek_BTln(a_{e^{-}}) \ + \ n_e e \upvarphi_{el} \nonumber \\ 
      &=\ 0 
\label{eq:half-final}
\end{align}

In the above expression $\upvarphi_{el}$ and $\upvarphi_{sol}$ correspond to the electrostatic potential in the electrode and the solution, respectively. Moreover, it has to be noted that in this formulation $z$ and $n_e$ are identical by definition. However, for clarity, at this point $z$ is used for the charge of the oxidized metal and $n_e$ for the number of electrons. Next, introducing the change in Gibbs free energy at the standard state $\Delta G^0$ (with $\Delta G^0 = \mu_M^0 - \mu_{M^{z+}}^0 - n_e\mu_{e^-}^0$), followed by a subsequent regrouping, yields:

\begin{equation}
 \Delta G^0 \ + \ k_BT ln \frac{a_{M}}{a_{M^{z+}} \ + \ {a_{e^{-}}^{n_e}}} \ + \ n_e e (\upvarphi_{el}-\upvarphi_{sol}) \ = \ 0
\end{equation}

It has to be noted that $\Delta G^0$ is directly related to the standard electrode potential ($E^0 = -\frac{\Delta G^0}{n_e e}$) and in principle still contains a temperature dependence. In other words, the usually applied standard electrode potential is strictly speaking only valid at standard conditions and actually decreases with temperature ($\frac{\partial E^0}{\partial T}$ is typically of the order of 1 mV/K).
Furthermore, by identifying the potential difference $\upvarphi_{el}-\upvarphi_{sol}$ with $\Delta \upvarphi$, finally the Nernst equation for the corresponding half cell is obtained:

\begin{equation}
 \Delta \upvarphi \ = \ E^0 - \frac{k_BT}{n_e e} ln \frac{a_{M}}{a_{M^{z+}} \ {a_{e^{-}}^{n_e}}} \ = \ -\frac{\Delta G_R}{n_e e}
 \label{eqn:Nernst-half-cell}
\end{equation}

In the last step, $\Delta G_R$ represents the Gibbs free energy of reaction for the electrode. The above derived Nernst equation for a half cell can now easily be extended to a general formulation for the full cell. Clearly, the open circuit voltage (OCV) of any battery is determined by the respective half--cell reactions on the anode and cathode side and thus by the resulting overall change in Gibbs free energy during the electrochemical reactions in the full cell, thus yielding:
\begin{align}
 U_{OCV} &= \Delta \upvarphi_{cathode} \ - \ \Delta \upvarphi_{anode} \nonumber\\ 
           &= \Delta E^0_{cell} \ - \ \frac{k_BT}{n_e e} ln \prod_i a_i^{v_i}
\label{eqn:Nernst-full-cell}
 \end{align}

In practice, at low concentration the activity coefficients $a_i$ can usually be replaced by the concentration of the respective element. As will be discussed in the next paragraph, a computational treatment of a half cell is possible by making use of the computational hydrogen electrode concept. Afterwards, this formal thermodynamic treatment will be applied to the case of Li-- and post--Li--ion type batteries, resulting in a significant simplification of the above equation. 
 
\subsubsection{Computational hydrogen electrode (CHE)}
\label{section:CHE}

Before continuing with the case of Li--ion type batteries, we focus on the above derived general case and line out how the reactions at the respective electrodes can be described computationally. For this purpose, we again take the expression of the Nernst equation for the half cell as given in Eq.~(\ref{eqn:Nernst-half-cell}). As already introduced above, this formalism contains the electrochemical potentials of the solvated ions in the respective half cell. Therefore, a direct computation would be extremely expensive, as a proper modelling of solvation free energies in principle necessitates explicit ensemble averages and thermodynamic integration schemes~\cite{Gross2019}.
Fortunately, a very elegant concept to circumvent this issue, which is frequently used in surface science, does exist: The computational hydrogen electrode (CHE). Indeed, the CHE allows to investigate half cells within a grand--canonical approach, yet, without increased computational cost~\cite{Norskov2004, Gross2021}. 

The CHE concept makes use of the fact that at standard conditions, which define the standard hydrogen electrode (SHE), hydrogen in the gas phase and protons in solution are in equilibrium, meaning $\tilde{\mu}_{{H^{+}}_{(aq)}} \ + \ \tilde{\mu}_{e^{-}} \ = \ \frac{1}{2} \mu_{H_2}$. Furthermore, the dependence of the SHE on the electrode potential and pH value (or concentration) is well--known. As a consequence, instead of addressing the solvated proton, the computationally accessible hydrogen molecule in the gas phase can be used as a reference~\cite{Norskov2004,Gossenberger2016, Gross2021}. Applying this approach for the case of hydrogen, one arrives at the following expression with respect to the electrochemical potential of proton and electron in solution:
\begin{equation}
\tilde{\mu}_{{H^{+}}_{(aq)}} \ + \ \tilde{\mu}_{e^{-}} \ = \ \frac{1}{2} \mu_{H_2} \ - \ e U_{SHE} \ - \ k_BT ln(10) pH
\end{equation}
Here, $U_{SHE}$ stands for the electrode potential with respect to the standard hydrogen electrode potential. 
An analogous extension of this concept to other redox couples is then easily possible~\cite{Gossenberger2015, Gossenberger2016}:
\begin{multline}
\tilde{\mu}_{{Me^{z+}}_{(aq)}} \ + \ z\tilde{\mu}_{{e^{-}}} \ = \ \mu_{Me} \ - \ ze (U_{SHE} - U_0) \\ - \ k_BT ln(a_{Me^{z+}})
\end{multline}
with $U_0$ the reduction potential of the $Me/Me^{z+}$ couple with respect to the SHE scale and the pH replaced by the activity. 
With these expressions for the electrochemical potentials of the charged ions in solution, the half--cell reaction in Eq.~(\ref{eqn:Nernst-half-cell}) can be determined without explicit calculation of the solvated species.

In practice, this grand--canonical approach is often applied to determine phase diagrams as a function of the electrochemical environment -- again, this is in particular used in surface science to investigate the most stable surface coverage for given conditions\,\cite{Magnussen2019}. For this purpose, the dependence of the electrochemical potential on temperature, concentration and applied potential can formally be combined to a single term $\Delta \tilde{\mu}$, which is obtained by subtracting the total energy of the bulk phase from the electrochemical potential of solvated proton and electron. This normalization step corresponds to the assumption that the bulk phase free energy is independent of temperature and electrochemical environment and hence can be approximated by the total energy, which for instance is directly accessible by density functional theory (DFT) calculations.
For the proton this then results in the following equation:
\begin{align}
\Delta \tilde{\mu}_{H^{+}}(T, p, U) \ = \ \tilde{\mu}_{{H^{+}}_{(aq)}}(T, p, U) \ + \ \tilde{\mu}_{e^{-}} \ - \ \frac{1}{2} E_{H_2} \nonumber\\
\ \approx \ -e U_{SHE} \ - \ k_BT ln(10) pH
\end{align}

For the case of metal species the same line of thought yields:

\begin{align}
\Delta \tilde{\mu}_{Me^{z+}}(T, p, U) \ = \ \tilde{\mu}_{{Me^{z+}}_{(aq)}}(T, p, U) \ + \ z\tilde{\mu}_{e^{-}} \ - \ E_{Me} \nonumber \\ 
\ \approx \ -ze(U_{SHE}-U_0) \ - \  k_BT ln(a_{Me^{z+}})
\end{align}

\begin{figure}[t]
\center\includegraphics[angle=0,width=0.95\columnwidth]{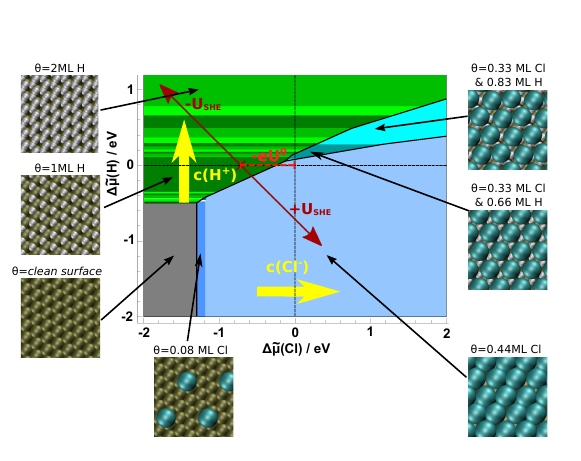}
\caption{Schematic respresentation of a surface phase diagram with respect to the normalized electrochemical potential for the co--adsorption of chlorine and hydrogen on a Pt (111) surface. Reprinted from Gossenberger et al.~\cite{Gossenberger2016} with permission from Elsevier.}
\label{fig:CHE}
\end{figure}

This approach is well--known for accessing the phase diagrams of electrode surfaces~\cite{Peterson2010, Gossenberger2016, Gossenberger2020, Didar2021}, where the change in Gibbs free energy, $\Delta G$, due to the adsorption of a certain species on an electrode surface is typically expressed as a function of the normalized electrochemical potential $\Delta \tilde{\mu}$ of the involved species:
\begin{equation}
\Delta G \ \approx \ E_{ads} \ - \ \sum_i n_i\Delta \tilde{\mu}_i(T, p, U)
\end{equation}
Here, $E_{ads}$ corresponds to the energy difference between the surface with adsorbed species on the one hand and the clean surface and the bulk phase of the adsorbates on the other. Note that for the discussion of electrode surfaces the change in Gibbs free energy is typically additionally normalized to the surface area ($\Delta \gamma = \Delta G/A$). The most stable phase for given conditions is then the one with the lowest $\Delta \gamma$, thus translating in a phase diagram~\cite{Gossenberger2016} as depicted in Fig.~\ref{fig:CHE}.

This formalism can now easily be transferred to investigate the stability of intercalation compounds -- obviously there is a close analogy between the case of adsorption on a surface and an intercalation process -- under given electrochemical conditions. 
For this purpose, the adsorption energy simply has to be replaced by the insertion energy, which analogously becomes the energy of formation of the intercalation phase with respect to the pristine electrode material and the bulk phase of the charge carrier. This now allows to evaluate the stable bulk phases at given electrochemical conditions, exactly as in the case of a surface. Moreover, for a given electrode reaction the CHE approach can then also be used to determine the corresponding voltage of the half cell with respect to the SHE.

\subsection{Alkali metal ion batteries}

The expression for the respective half cell potentials as derived in Eq.~(\ref{eqn:Nernst-half-cell}) is fairly complicated and would, without the work-around introduced in the previous paragraph, be highly demanding to access computationally. However, in case of alkali metal ion batteries such as LIBs, we are even in a more fortunate situation as only one species is involved in the overall process. In fact, the effective thermodynamic process that occurs during discharge is the transfer of a Li atom from anode to cathode, such that the actual solvation (desolvation) process of Li$^+$ ions at the anode (cathode) and its energy gain (cost) do not have to be considered in the total energy balance. In other words, for determining the energy gain per transferred electron (\textit{i.e.} the voltage) of a LIB, the half--cell reactions do not have to be explicitly considered. The corresponding open circuit voltage can instead directly be obtained in terms of the overall change in Gibbs free energy between the respective states of anode and cathode material:

\begin{equation}
 U_{OCV} = -\frac{\Delta G}{n_e e}
 \label{eq:OCV}
\end{equation}

In Eq.~(\ref{eq:OCV}), $\Delta G$ corresponds to the change in the Gibbs energy, whereas $U_{OCV}$ is the resulting open circuit voltage. $n_e$ is the number of electrons that is transferred between anode and cathode, while $e$ denotes the elementary charge. When $\Delta G$ is expressed in eV, then $U_{OCV}$ in volts can be very conveniently determined by just dividing this value by the number of transfered elctrons $n_e$.
If the Gibbs free energy and the charge transfer are given per mole, Eq.~(\ref{eq:OCV}) transforms into the frequently used form:
\begin{equation}
 U_{OCV} = -\frac{\Delta G}{z F} ,
 \label{eq:OCV2}
\end{equation}
where $F$ corresponds to the Faraday constant, whereas $z$ denotes the valency of the charge carrier, \textit{i.e.} 1 in the case of alkali metal atoms and 2 in the case of alkaline earth atoms.

In the following, the determination of the OCV for an archetype LIB with a graphite intercalation compound (GIC) as anode and a transition metal (TM) oxide cathode will be exemplified. For this purpose, the following full cell reaction has to be considered:
\begin{equation}
 \mathrm{Li}_{x_1+\Delta x}\mathrm{C}_{6} \ + \ \mathrm{Li}_{x_2}\mathrm{TMO}_2 \rightarrow \mathrm{Li}_{x_2+\Delta x}\mathrm{TMO}_2 \ + \ \mathrm{Li}_{x_1}\mathrm{C}_{6}
\end{equation}

Here, $\Delta x > 0$ is assumed, such that the reaction direction represents the discharge of the battery, \textit{i.e.}, Li deintercalates at the GIC anode and is transferred to the TM oxide cathode. As discussed above, the OCV for this cell reaction is then obtained from the corresponding overall difference in Gibbs free energy:
\begin{multline}
 \Delta G \ = \ G(\mathrm{Li}_{x_2+\Delta x}\mathrm{TMO}_2) \ + \ G(\mathrm{Li}_{x_1}\mathrm{C}_{6}) \\ \ - \ [G(\mathrm{Li}_{x_2}\mathrm{TMO}_2) \ + \ G(\mathrm{Li}_{x_1+\Delta x}\mathrm{C}_6)]
 \label{eq:redox}
\end{multline}

As already pointed out before, this simply means that we consider the free energy gain for moving a Li atom from the anode side to the cathode side. In practice, the free energy expressions are typically approximated by the total energies, which in turn can easily be obtained from DFT calculations, thus yielding:

\begin{multline}
  \Delta G \approx \Delta E \ =  \ E_{tot}(\mathrm{Li}_{x_2+\Delta x}\mathrm{TMO}_2)  \ + \ E_{tot}(\mathrm{Li}_{x_1}\mathrm{C}_{6}) \\ \ - \ [E_{tot}(\mathrm{Li}_{x_2}\mathrm{TMO}_2) \ + \ E_{tot}(\mathrm{Li}_{x_1+\Delta x}\mathrm{C}_6)]
 \label{eq:GtoE}
\end{multline}

This assumption is justified by the fact that the total energy is the dominant contribution to the Gibbs free energy, while \textit{e.g.}, $pV$ (for a solid) as well as vibrational and configurational entropy are typically rather small. In fact, the configurational entropy of a binary solid solution amounts to less than $\sim$ 20 meV/atom at room temperature, whereas vibrational entropy typically lies roughly in the same range. Moreover, only the entropy differences between the different states are actually of importance, which means that a large part of the entropic contribution typically cancels out, thus resulting in overall errors in the order of 0.01 V~\cite{Rogal2007,Meng2009}.
Finally, the voltage at a given state of charge (\textit{i.e.}, between $x_1 + \Delta x $ and $x_1$) can be directly obtained via:

\begin{equation}
 U(x_1+\Delta x,x_1)= -\frac{\Delta E}{e (\Delta x)}
 \label{eq:Voltage}
\end{equation}

Remember that the discharge process formally corresponds to the transfer of neutral Li atoms from anode to cathode. Under the assumption of constant temperature and pressure, the change in Gibbs free energy can also be expressed by the respective Li chemical potential, $\mu_{Li} = \left(\frac{\partial{G_{Li}}}{\partial{n_{Li}}}\right)_{P,T}$:
\begin{equation}
 U_{OCV} \ = \ -\frac{\Delta G}{n_e e} \ = \ -\frac{\mu_{Li}^{cathode} \ - \ \mu_{Li}^{anode}}{e},
\label{eqn:LIB}
 \end{equation}

\noindent with $\mu_{Li}^{cathode}$ and  $\mu_{Li}^{anode}$ the Li chemical potential at the cathode and the anode side, respectively.

\begin{figure}[t]
\center\includegraphics[angle=0,width=0.95\columnwidth]{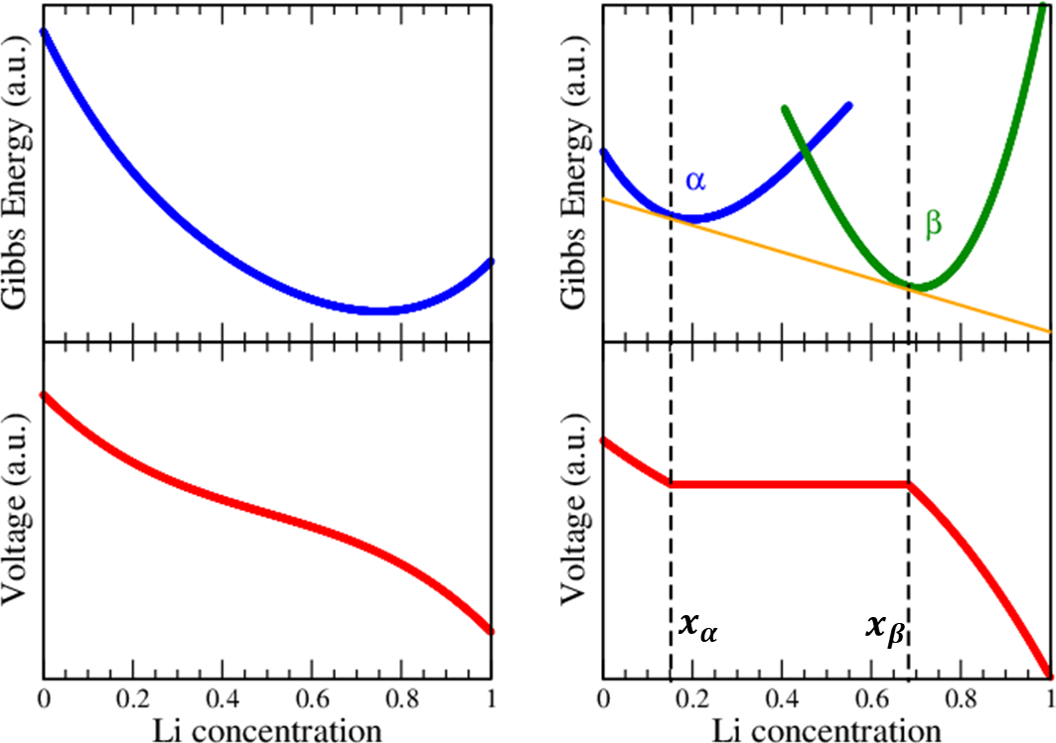}
	\caption{Gibbs free energy of two different cathode materials and the corresponding voltage profiles with respect to the chemical composition (Li$_x$M) for the prototypical cases of solid solution behavior (left) and phase separation into two phase $\alpha$ and $\beta$ (right). The straight yellow line corresponds to the common tangent construction (see reference~\cite{vanderVen2013} for similar discussion).}
\label{fig:phase-diagram}
\end{figure}

\subsubsection{Thermodynamic interpretation of charge/discharge curves}
An experimental standard method for characterizing the performance of anode and cathode materials is the determination of the charge--discharge profile, \textit{i.e.}, the evolution of the voltage with respect to the state of charge. While the resulting voltage profiles are frequently used to qualitatively describe the respective materials, the direct relation of the charge--discharge curve to the underlying phase diagram is not always considered. To clarify this point, the Gibbs energy and the corresponding voltage profile for two prototypical cases, namely phase separation and solid solution behavior, are depicted in Fig.~\ref{fig:phase-diagram}.

Indeed, the voltage profile of a cathode material is directly related to the Gibbs free energy of the (meta--) stable phases (with different Li concentrations) the system passes through during discharge (or charge). Hence, the phase diagram determines which phases have to be considered and where phase transition are to be expected. 
As derived in Eq.~(\ref{eqn:LIB}), the voltage profile of a cathode material with respect to a given anode can then be obtained from their difference in chemical potential. For a Li--metal anode the corresponding Li chemical potential $\mu_{Li}^{anode}$ is constant, while the Li chemical potential at the cathode is obtained from the derivative of its Gibbs free energy with respect to the concentration. Hence, as is evident from Eq.~(\ref{eqn:LIB}), this derivative directly determines the shape of the resulting voltage profile. From Fig.~\ref{fig:phase-diagram} it becomes clear that a slopy potential is the signature of a solid solution process, whereas a plateau in the charge--discharge curve is directly related to a phase separation~\cite{vanderVen2013}. The solid solution process, as depicted in the left panel of Fig.~\ref{fig:phase-diagram}, means that the system can accomodate each Li concentration without significant structural changes (phase transition). In the right panle of Fig.~\ref{fig:phase-diagram}, on the other hand, the phase separation into the two phases $\alpha$ and $\beta$ is depicted. The common tangent construction in the top panel (yellow line), shows that between the two limiting concentrations, indicated by the dashed lines, the coexistence of the $\alpha$ and the $\beta$ phase, with Li concentrations $x_{\alpha}$ and $x_{\beta}$, will always be the energetically most favorable situation. Hence, a phase separation occurs, which results in a voltage plateau in the charge--discharge curve, as depicted in the bottom panel.

\begin{figure}[t]
\center\includegraphics[angle=0,width=0.95\columnwidth]{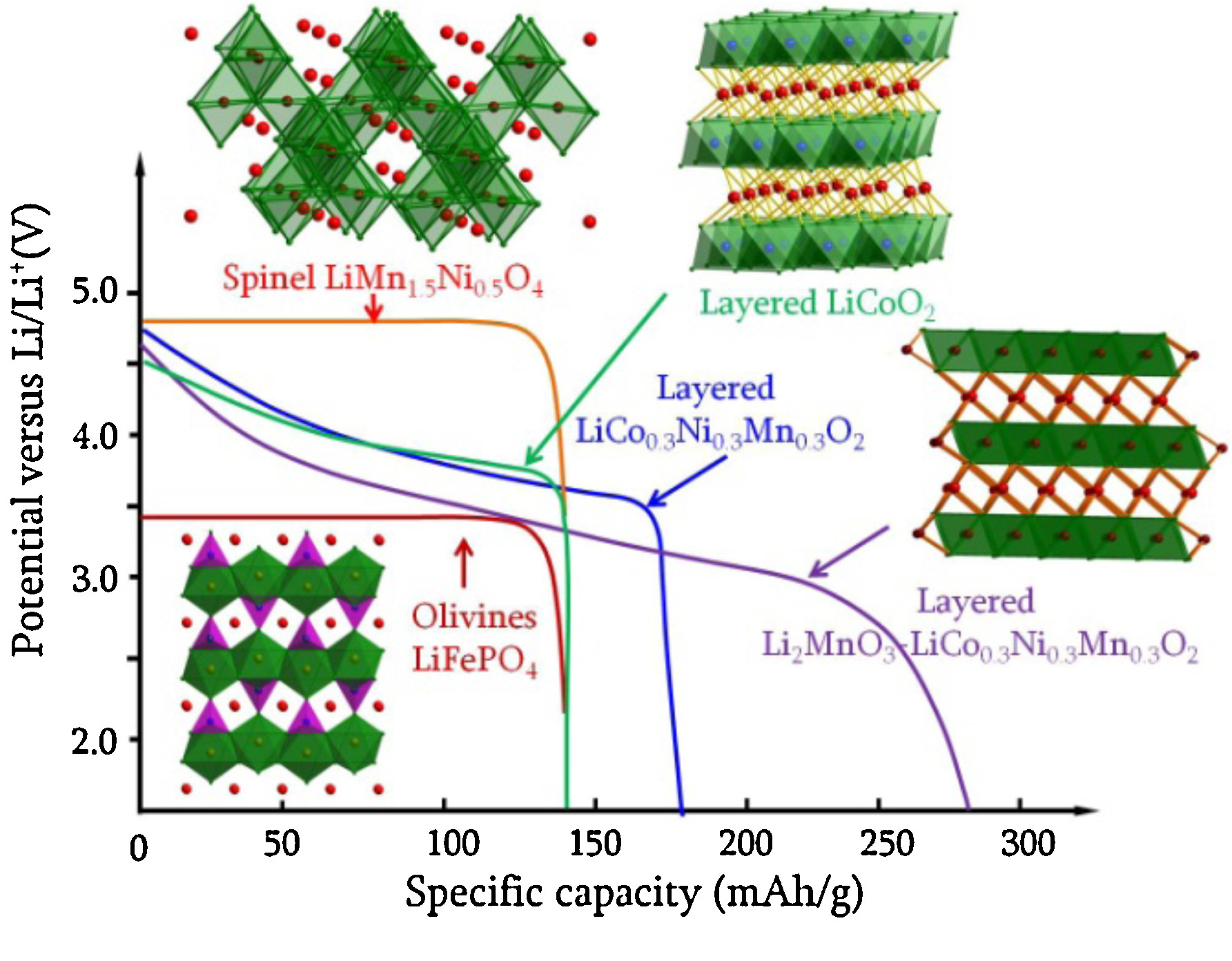}
\caption{Voltage profile for solid solution and phase separation behaviour for several characteristic cathode materials. Reprinted from Liu et al.~\cite{Liu2016b}. With permission from Elsevier.}
\label{fig:profiles}
\end{figure}

In Fig.~\ref{fig:profiles}, experimentally determined discharge curves for several different cathode materials are depicted~\cite{Liu2016b}. The voltage profiles of Li$_x$CoO$_2$ and the different layered NMCs are prototypical examples for a solid solution behavior, whereas the spinel and olivine type compounds show the characteristic step profile with a pronounced plateau that indicates different phases and thus phase separation. Finally, it has to be pointed out that for real electrodes discharge and charge profile do not fall on top of each other but are separated by a certain offset, frequently referred to as polarization. This is a consequence of kinetic limitations due to, for instance, increased diffusion barriers.

\subsubsection{Further insights into the electrochemical potential}
As already introduced in Eq.~(\ref{eqn:LIB}), for the transfer of a Li--ion (and the corresponding electron) from anode to cathode the OCV is fully determined by the difference in the respective Li chemical potential on the cathode and the anode side.

Starting from this expression, additional electrochemical considerations can be made. In fact, the Li chemical potential can be denoted as the sum of the electrochemical potentials of a Li$^+$ ion and the corresponding electron, as already briefly introduced in the thermodynamic derivation of the Nernst equation:
\begin{equation}
 \mu_{\mathrm{Li}} \ = \ \tilde\mu_{\mathrm{Li}^{+}} \ + \ \tilde\mu_{e^{-}}
\end{equation}

\begin{figure*}[t]
 \includegraphics[angle=0,width=0.4\linewidth]{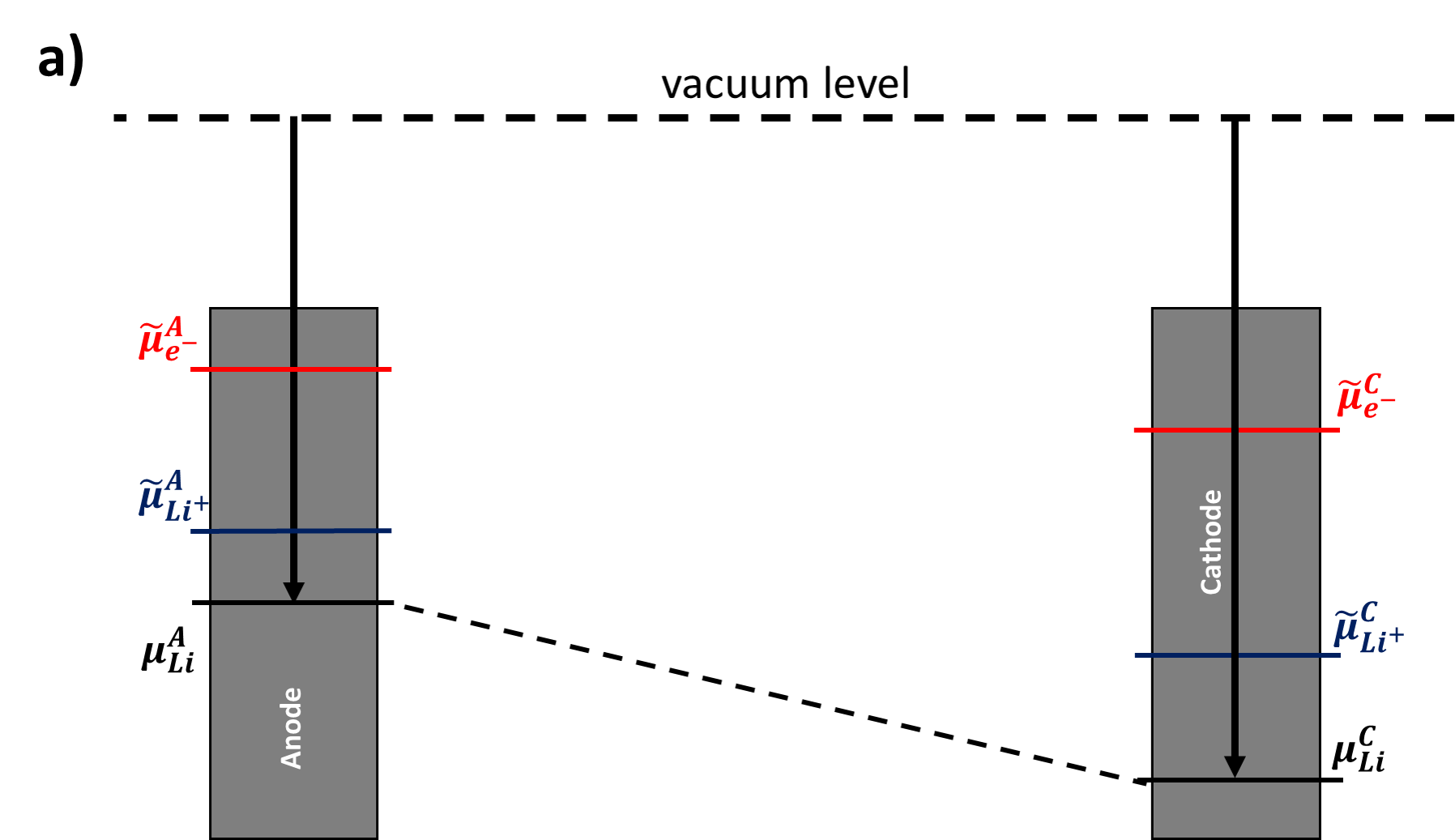}\hspace{2cm}
 \includegraphics[angle=0,width=0.4\linewidth]{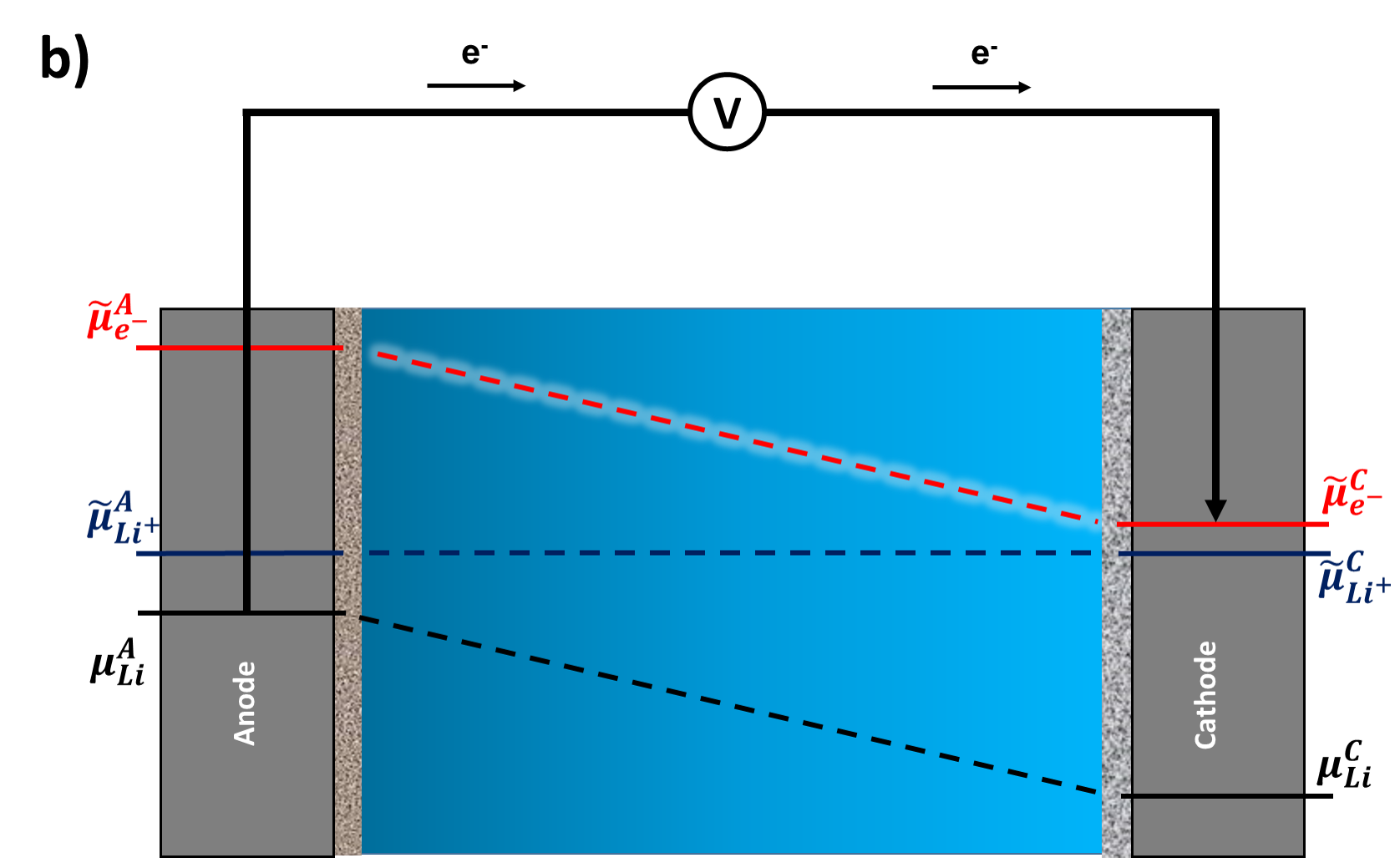}
 \caption{Chemical potential of Li as well as electrochemical potentials of Li$^{+}$ and e$^{-}$ for anode and cathode material a) before and b) after introduction of an electrolyte under open circuit conditions. For the electrode immersed in electrolyte, it has to be noted that $\tilde{\mu}_{Li^+}$ is equal in anode, cathode and electrolyte. The dashed lines in the case of $\tilde{\mu}_{Li^+}$ and $\tilde{\mu}_{e^-}$ are just a guide for the eye and should not be understood as realistic potential curves. Partially redrawn from~\cite{Gross2019}.}
 \label{fig:potential}
\end{figure*}

Now, to gain further insight, the exemplary case of two electrodes in vacuum that are not connected will be discussed. For this situation, the respective potentials have to be determined relative to the vacuum level, as depicted in Fig.~\ref{fig:potential}. Next, it is assumed that this anode and cathode configuration is immersed in an electrolyte. Keeping in mind that the electrolyte is able to shuttle Li$^+$ ions between the electrodes to establish an equilibrium distribution, the Li$^+$ electrochemical potential in the anode, cathode and electrolyte have to be identical. To put it differently, if $\tilde\mu_{\mathrm{Li}^{+}}$ in electrode and electrolyte was different, the Li$^+$ ions in the electrolyte would compensate the differences by adapting their concentration profile until $\tilde\mu_{\mathrm{Li}^{+}}$ becomes constant throughout the system~\cite{Gross2019}. Interestingly, the fact that under these equilibrium conditions $\tilde\mu_{\mathrm{Li}^{+}}$ is the same for anode and cathode (and electrolyte) means that, for the electrolyte containing system, the difference in electron electrochemical potential at the anode and cathode side actually determines $U_{OCV}$:
\begin{equation}
 U_{OCV} \ = \ -\frac{\mu_{Li}^{cathode} \ - \ \mu_{Li}^{anode}}{e} \ = \ -\frac{\tilde\mu_{e^-}^{cathode} \ - \ \tilde\mu_{e^-}^{anode}}{e}
\end{equation}

The relationship between $\mu_\mathrm{Li}$, $\tilde\mu_{\mathrm{Li}^{+}}$ and $\tilde\mu_{e^{-}}$ at the anode and cathode side in vacuum and in presence of the electrolyte are schematically depicted in Fig~\ref{fig:potential}. Considering the full battery cell with immersed electrodes -- under open circuit conditions -- this also means that the electronic levels have to adjust as compared to the situation in vacuum. As just derived, this must happen in such a way that the difference in the electron electrochemical potential corresponds to the OCV~\cite{Gross2019}, which can be inferred by comparing the schematic drawings in Fig.~\ref{fig:potential}.
Of course, now the question arises how this adjustment can be understood on a microscopic scale. In fact, a convincing explanation relies on interpreting the level alignment in terms of the electric double layer formation at the anode and cathode side in the presence of an electrolyte. Within the double layers, electric fields are created, which equate the electrochemical potentials of the Li$^+$ ions. Similarly, the electrons are subject to the same field, but due to their opposite charge their electron electrochemical potentials in cathode and anode shift in such a way that the 
final difference amounts to the open--circuit voltage. Consequently, with regard to their respective positions for the single electrodes in vacuum, the electrochemical potential of Li$^+$ ions and electrons would shift in opposite directions~\cite{Gross2019}.

\begin{figure}[b]
 \center\includegraphics[angle=0,width=0.98\columnwidth]{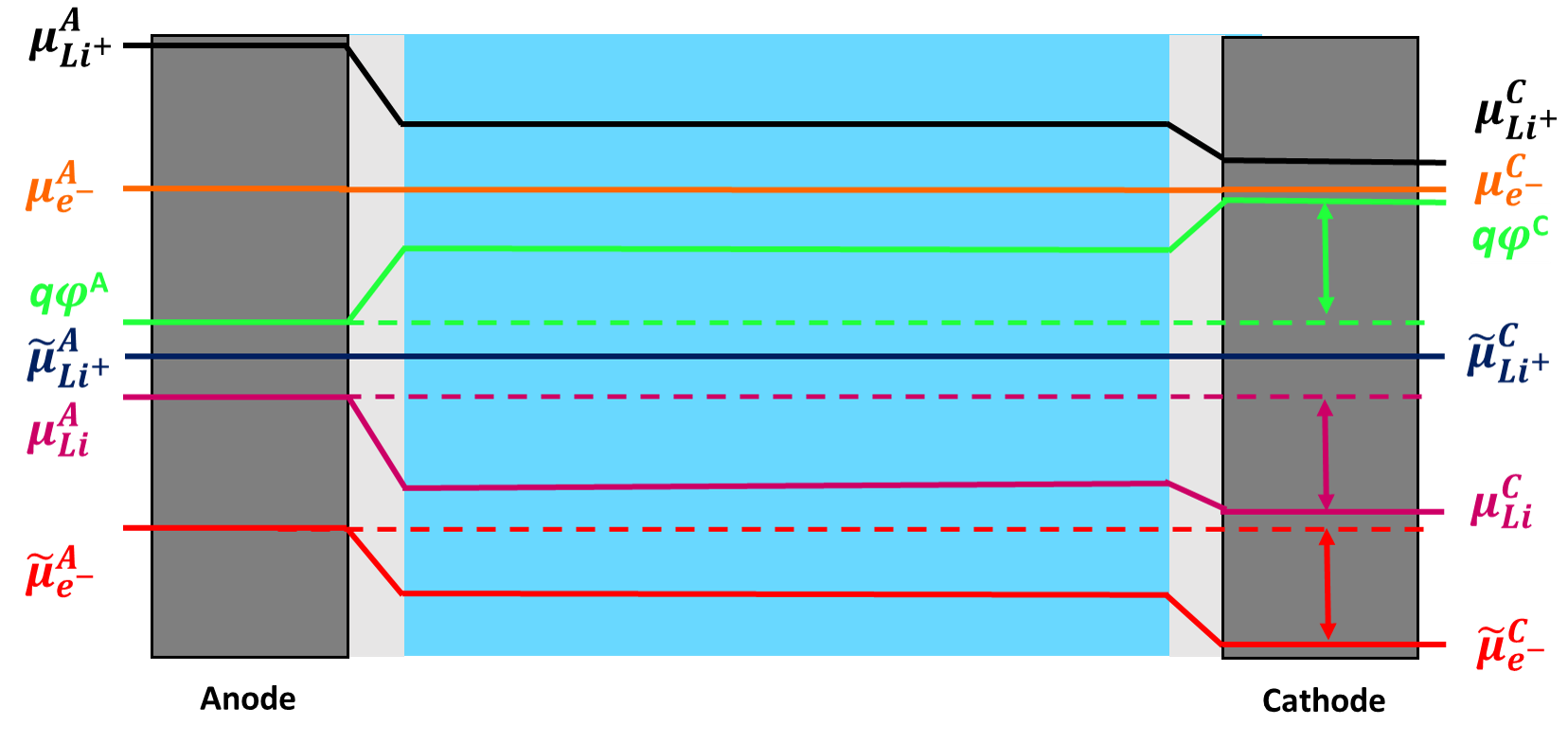}
 \caption{Potential curves between anode and cathode for a battery with liquid electrolyte at open circuit conditions. While the linear potential changes in the light gray areas represent the electric double layer, it has to be noted that the indicated potential drop/increase is a very crude approximation and only serves for illustrative purposes. In particular, it should be kept in mind that in the bulk of anode, cathode and electrolyte the potentials are constant. Potential changes and hence electric fields are restricted to the double layer.}
 \label{fig:potential-schematic}
\end{figure}

Moreover, it should be noted that in equilibrium, the potential $\upvarphi$ and hence also the chemical potential $\mu_{Li^+}$ of the Li$^+$ ions is constant throughout the bulk electrolyte. For further clarification, the relations between the different chemical and electrochemical potentials and their spatial dependence are again schematically depicted in Fig.~\ref{fig:potential-schematic}. Hence, for open circuit condition there are neither electric fields in the bulk of the electrodes nor in the bulk electrolyte. 
Electric fields have to be present, yet, are limited to the interfaces and result from changes in the electrostatic potential. These changes are qualitatively indicated by a slope in Fig.~\ref{fig:potential-schematic}, which, however, has to be understood as a schematic illustration. The real evolution of the double layer potential takes a more complex form that goes even far beyond frequently used descriptions like the Stern model~\cite{Sakong2018,Gross2019}.

\subsubsection{Redox concepts}

As already discussed in the introduction, the voltage of a battery is determined by both anode and cathode material (see Fig.~\ref{fig:Landi}). In fact, in an ideal battery the chemical potential of Li in the anode should lie as close as possible to the potential of the Li/Li$^+$ couple \textit{i.e.}, close to the chemical potential of Li metal as this corresponds to the thermodynamical stable form of solid Li. However, for Li metal anodes, Li plating can occur, going along with the risk of dendrite growth and battery failure~\cite{Jaeckle2018, Wang2020}. Hence, Li metal is nowaydays only considered for solid state batteries, while standard LIBs usually rely on graphite based anodes. In a simplified picture, the chemical potential of the cathode on the other hand is predominantly determined by the active redox couple in the respective compound. This means that \textit{e.g.}, in a LiCoO$_2$ cathode the observed potential profile is a consequence of the formal change in oxidation state of the Co$^{3+}$/Co$^{4+}$ redox couple under delithiation/lithiation (charging/discharging) of the electrode:

\begin{equation}
 \mathrm{LiCo}^{+3}\mathrm{O}_2 \ \rightleftharpoons \ \mathrm{Li} \ + \ \mathrm{Co}^{4+}\mathrm{O}_2
 \label{eqn:redox}
\end{equation}

It should be noted that Eq.~(\ref{eqn:redox}) assumes full delithiation, a scenario which is rarely possible for realistic cathode materials. In fact, for the case of layered oxides such as LiCoO$_2$, the repulsion between the transition metal oxide layers is getting strongly increased under delithiation and finally results in a collapse of the structure. The exemplary case of LiCoO$_2$ actually allows only for about half of its theoretical capacity to be exploited (\textit{i.e.}, 0.5 Li/f.u.), before the structural integrity is affected~\cite{Reimers1992}.

\begin{figure}[t]
 \center\includegraphics[angle=0,width=0.7\columnwidth]{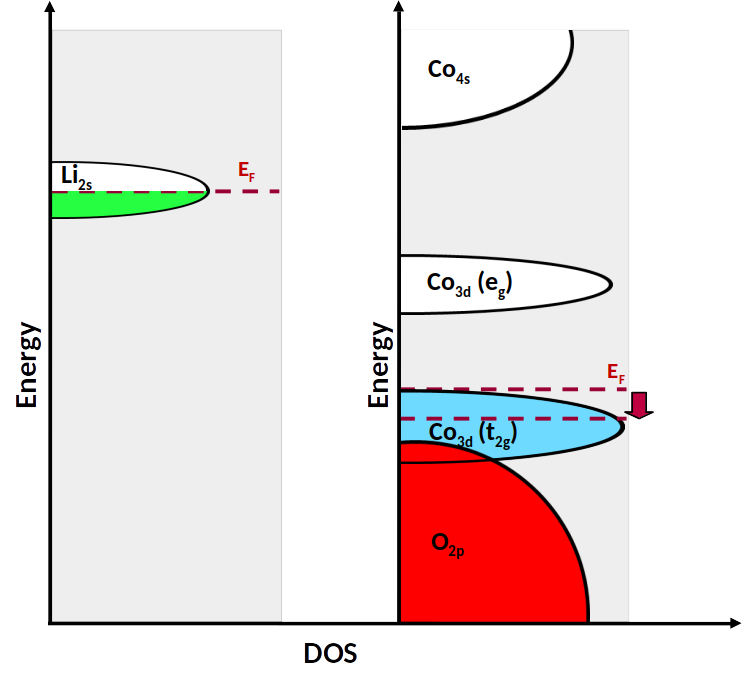}
 \caption{Qualitative picture of redox processes, exemplified with respect to the electronic density of states (DOS) in a LiCoO$_2$ cathode under operation~\cite{Goodenough2013,Hausbrand2015,Gao2016}. The anode Fermi level of the Li metal anode is unchanged (left), while the cathode Fermi level shifts down during charging (delithiation) as indicated by the red arrow between the two dashed lines (right).  }
 \label{fig:orbitals}
\end{figure}

In Fig.~\ref{fig:orbitals}, a schematic diagram of the electronic structure of a LiCoO$_2$ cathode in combination with a Li--metal anode is depicted, corresponding to the microscopic picture of the macroscopic situation discussed in Fig. \ref{fig:potential}, with the OCV determined by the difference in the electron electrochemical potential. For this particular case, the Co 3$d$ states are located above the oxygen $2p$ states and are hence the states that are involved in the redox process. Thus, when the battery is charged (delithiated), Co is oxidized from 3+ to 4+ and consequently electrons are removed from the 3$d$ states.

\begin{figure}[b]
 \includegraphics[angle=270,width=0.9\columnwidth]{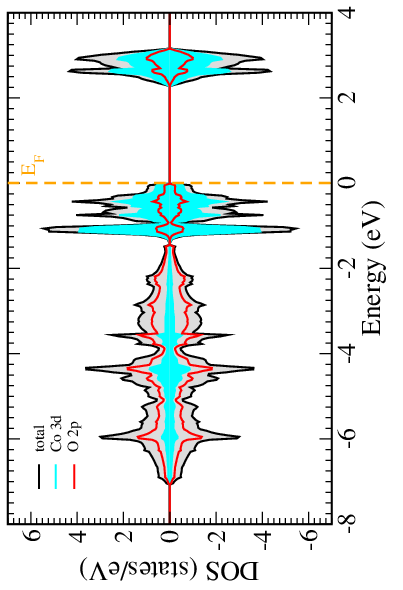}
 \caption{Electronic density of states (DOS) for LiCoO$_2$, as obtained from DFT calculations that have been performed for this review, applying the SCAN Meta--GGA exchange--correlation functional. The partial DOS for oxygen \textit{p}-- and Co \textit{d}--states are depicted in red and cyan, the total DOS is shown in black.}
 \label{fig:DFT-DOS}
\end{figure}

As already mentioned above, this oxidation state picture, however, corresponds to a simplified view. In fact, the description of the redox process as adding (removing) electrons to (from) unaltered Co 3$d$ states holds only for a so--called rigid band model, which in our case assumes that both the electronic and the crystal structure of the LiCoO$_2$ cathode are unaltered by the insertion/removal of a Li atom~\cite{Meng2009}.
While this description allows to gain an intuitive and qualitatively correct picture of the underlying redox process, it is of limited validity~\cite{Meng2009,Hausbrand2015}. This becomes evident when comparing the schematic figure and the DFT calculated density of states in Figs.~\ref{fig:orbitals} and~\ref{fig:DFT-DOS}. The DFT calculations yield a qualitatively similar picture as the schematic drawing in Fig.~\ref{fig:orbitals}, yet, the details are different. In fact, the partial DOS indicates that indeed the Co 3$d$ states are dominant below the Fermi level, however, there is also some hybridization with the oxygen 2$p$ states. Hence, the delithiation will also affect the oxygen states as opposed to the simplified redox picture discussed above. Furthermore, the removal of Li atoms will also affect the details of the crystal structure, which in turn will also have an impact on the electronic structure.

\begin{figure}[t]
\includegraphics[angle=0,width=0.9\columnwidth]{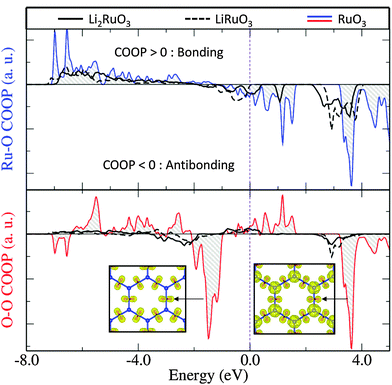}
 \caption{Oxo to peroxo transformation upon delithiation of Li$_2$RuO$_3$ (2O$^{2-}$/(O$_2$)$^{n-}$). The computed COOPs for Ru--O and O--O bonds are depicted. The increasing O--O COOP amplitude indicates O--O bond formation under oxidation of Li$_{2-x}$RuO$_3$. The corresponding electron densities of the bands with the strongest anti-bonding O--O bonds show a polarization of the O 2$p$--orbitals which results in the formation of $\sigma$-type O--O bonds\,\cite{Saubanere2016}. With permission of the Royal Society of Chemistry.}
 \label{fig:COOP}
\end{figure}

Obviously, in non--ideal redox systems it comes not as a surprise that species other than the transition metals may participate in a redox process. In this regard, anionic redox is an often discussed scenario, offering new pathways for increasing the capacity of cathode materials~\cite{Yahia2019,Chang2020}. In transition metal oxides, the case of anionic redox typically occurs when the TM 3$d$ states are located below a larger fraction of the oxygen 2$p$ states. In this case the anion \textit{p}--band will also be strongly involved in the redox process, which in turn may have negative consequences such as the formation and release of oxygen gas, hence resulting in the degradation of the electrode~\cite{Seo2016,Okubo2017,Hausbrand2015}. On the other hand, if reversible anionic redox is possible without deteriorating the electrode stability, it allows for further Li extraction and therefore gives access to additional capacity~\cite{Saha2019,Saubanere2016,Sathiya2013,Yahia2019}. Thus, an understanding of anionic redox and how to exploit it is a highly interesting question with respect to improving the capacity of cathode materials.

\begin{figure}
\includegraphics[angle=0,width=0.98\columnwidth]{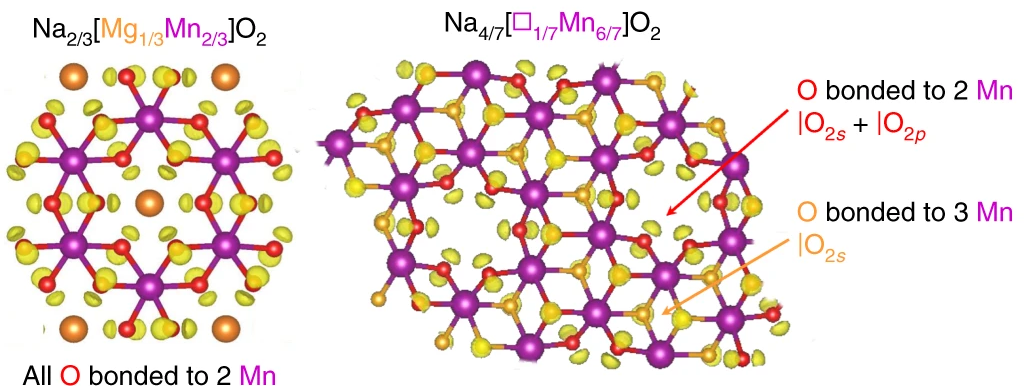}
 \caption{Electron localization function (ELF), as obtained from DFT calculations on Na$_{2/3}$Mg$_{1/3}$Mn$_{2/3}$O$_2$, showing one O \textit{2s} and one O 2$p$ oxygen lone pair per oxygen. For the case of Na$_{4/7}$Mn$_{6/7}$O$_2$, one and two oxygen lone pairs per oxygen occur in the Mn-rich (orange) and Mn-poor (red) oxygen environments, respectively~\cite{Yahia2019}. Reprinted from this reference with permission from Springer Nature.}
 \label{fig:ELF}
\end{figure}

For several oxide cathode materials it has been suggested that reversible anionic redox is only possible when the created peroxo--like $(O_2)^{n-}$ species interact with the transition metals through a reductive coupling mechanism. Such a scenario was recently demonstrated by DFT studies on the oxidation of Li$_2$RuO$_3$~\cite{Sathiya2013,Saubanere2016}. In fact, the investigation of changes in the charge distribution during delithiation in combination with a crystal orbital overlap population (COOP) analysis of the oxygen--oxygen bonding characteristics allowed to show a reductive coupling between peroxo--like (O$_2)^{n-}$ species and the transition metal (see Fig.~\ref{fig:COOP}). Extending these studies, more recently a unified picture of anionic redox was proposed~\cite{Yahia2019}. 
In this work, oxygen lone pairs have been detected by evaluating the electron localization function (ELF), finally resulting in suggesting the number of created holes per oxygen as the crucial parameter to quantify the reversibility of anionic redox, with a critical maximum value of 1/3 holes per oxygen atom (see Fig.~\ref{fig:ELF})~\cite{Yahia2019}. 

\subsection{The electrochemical stability window}
\begin{figure}[t]
 \center\includegraphics[angle=0,width=0.8\columnwidth]{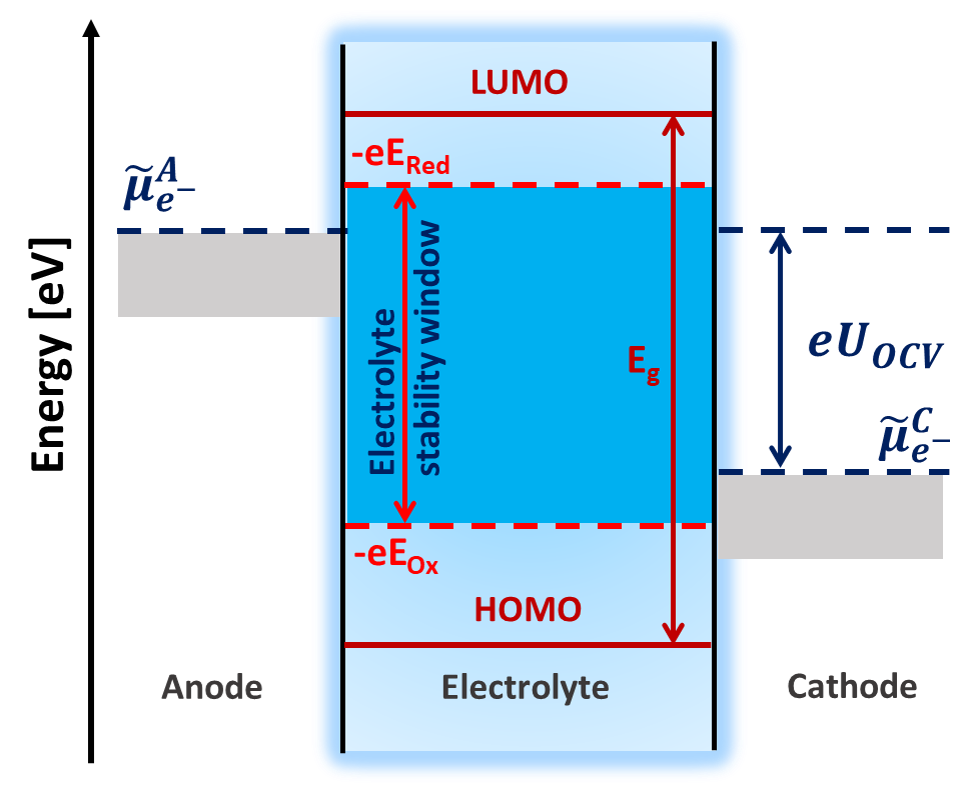}
 \caption{Schematic representation of the electrolyte stability window with respect to the chemical potential of an electron. HOMO and LUMO levels are shown for clarification (redrawn from~\cite{Peljo2018}).}
 \label{fig:window}
\end{figure}

Before closing this first part of the review, a still existing misconception with respect to electrolyte stability has to be discussed. Frequently, stability criteria for electrolytes are deduced by analyzing their HOMO and LUMO levels with respect to the Fermi level of anode and cathode, which is, however, a misleading concept~\cite{Peljo2018}. As will become clear immediately, it is not correct to identify the HOMO/LUMO gap with the stability window of the electrolyte. Instead, the stability of the electrolyte has to be determined with respect to the oxidation and reduction potential (\textit{i.e.} the thermodynamic stability), as indicated in Fig.~\ref{fig:window}~\cite{Peljo2018}. One of the most prominent examples, which makes it very clear that the HOMO/LUMO gap does not directly describe the stability window is certainly water. Whereas the HOMO/LUMO gap in water can be computed to extend to almost 9 eV~\cite{Chen2016}, in reasonable agreement with experimental data~\cite{Goulet1990,Winter2004}, the electrochemical stability window of water is tremendously smaller and amounts to the well--known 1.23 V. 
Therefore, it has to be pointed out once more that it is the stability with respect to oxidation and reduction reactions that determines the stability window of an electrolyte. Nevertheless, the HOMO/LUMO gap may possibly be interpreted as a kind of descriptor for the electrolyte stability by applying appropriate scaling relations~\cite{Borodin2013,Cheng2015}. 

\section{Computational Methods}  
\label{chapter:computational}


In the last decades density functional theory (DFT) has certainly evolved to one of the most widely used tools in computational material science. The increasing computer power in combination with the efficiency of DFT calculations allows for the accurate simulation of materials and processes on the atomic scale. 
While DFT calculations are typically performed for systems with a few 10 to 100 atoms, even simulations with 1000s of atoms are nowadays feasible with massively parallel codes running on supercomputers.
On the other hand, if processes on much larger length scales or for long time scales are of interest, usually more coarse grained models such as classical molecular dynamics, kinetic Monte Carlo (kMC) or cluster expansion based Monte Carlo approaches are the method of choice. 
However, often these coarse graining schemes are also based on data from DFT calculations, such that it seems justified to discuss the underlying principles of DFT in some detail.

\subsection{Density functional theory}

Most quantum mechanical descriptions of non--relativistic systems are simply based on the time--independent many--body Schrödinger equation:
\begin{equation}
H \ \Phi (\{\vec{R}\}, \{\vec{r}\}) \ = \ E \ \Phi (\{\vec{R}\}, \{\vec{r}\})
\label{SGL}
\end{equation}

Solving this eigenvalue problem allows to determine the ground state energy $E$ and the corresponding wave function $\Phi$. However, the many--body wave function $\Phi$ depends on the electron and core coordinates (${\vec{r}}$ and ${\vec{R}}$), thus making numerical solutions quite challenging already for small system sizes.
In fact, due to the high computational effort a joint quantum mechanical solution for both electronic and nuclear degrees of freedom is quickly intractable.
Fortunately, electrons and nuclei move on distinctly different time scales, such that the electronic and nuclear degrees of freedom can be decoupled, resulting in the famous Born--Oppenheimer approximation~\cite{Born1927}. This decoupling makes the nuclear coordinates $\{\vec{R}\}$ enter the remaining electronic Schrödinger equation as a set of parameters, resulting in the following expression for the electronic Schrödinger equation:
\begin{multline}
H_{el}(\{\vec{R}\}) \Psi(\{\vec{r}\}) \ = \ \left\{T_{e}  + \ V_{nn} \ + \ V_{ne} \ + \ V_{ee}  \right\}  \Psi(\{\vec{r}\})\\
= \ \left\{-\frac{\hbar^2}{2m}\nabla^2 \ + \ \frac{1}{2} \sum_{i,j}\frac{e^2}{\|\vec{r}_i - \vec{r}_j\|}\ - \ \sum_{i,J} \frac{Z_Je^2}{\|\vec{r}_i - \vec{R}_J\|} \right.\\
\left. \ + \ \frac{1}{2}\sum_{I,J} \frac{Z_IZ_Je^2}{\|\vec{R}_I - \vec{R}_J\|}\right\} \Psi(\{\vec{r}\}),
\label{eq:Hel}
\end{multline}  
with the kinetic energy $T_e$ and the potential energy terms, corresponding to the electrostatic interaction of the involved species, \textit{i.e.}, electron--electron interaction ($V_{ee})$), electron--nuclei interaction ($V_{ne}$) and nuclei--nuclei interaction ($V_{nn}$). 

\begin{figure}[b]
\includegraphics[angle=0,width=0.85\columnwidth]{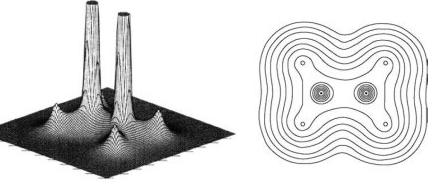}
\caption{Electron density of a C$_2$H$_4$ molecule. From the maxima in the electron density it is easy to extract the postion of the carbon and hydrogen atoms. Reprinted from~\cite{Piela}. With permission from Elsevier.}
\label{fig:edensity}
\end{figure}

Although the Born--Oppenheimer approximation indeed is a very successful approach, it should be noted that cases may exist for which its validity is not granted. While the remaining problem now has reduced to solving the electronic Schrödinger equation, this unfortunately still amounts to a non--trivial task, only being exactly solvable for the simplest situations. Consequently, already for treating small system sizes, further approximations are inevitable and this is where DFT sets in. 
The underlying principle of DFT is the Hohenberg--Kohn theorem~\cite{Hohenberg1964}, which states that the ground state of a system can uniquely be described by its electron density. 
At a first glance, this seems surprising, however, when looking at the electron density of a given system this correspondence agrees with our intuition. As exemplified in Fig.~\ref{fig:edensity}, the electron density of a molecule immediately allows to state where the nuclei are located. In addition, the derivative of the electron density contains information on the charge of a nucleus. Yet, only with the Hohenberg--Kohn theorem the formal and very elegant proof of such a correspondence was provided.
As a consequence, it is possible to express the Hamiltonian in Eq.~(\ref{eq:Hel}) as a functional of the electron density, 
which drastically reduces the dimensions and thus the complexity of the problem. Now, instead of determining the complex many--body wave function, only the electron density, being a function of the three spatial coordinates, needs to be considered. In principle, this problem could then be solved by determining the ground state energy using a variational approach with respect to the electron density, which would correspond to the so--called orbital--free DFT. Yet, due to its complicated and strongly non--local character, broadly applicable approximations to the kinetic energy functional $T_{e}[n]$ have, even up to now, remained elusive~\cite{Constantin2019}.
Hence an alternative approach has evolved. For the latter one, the ground state electron density is determined by mapping the Hamiltonian on a system of non--interacting electrons, which is designed in such a way that it exhibits the same ground state electron density as the original Hamiltonian. In this formulation, the electron density is then simply obtained from the single particle states of the non--interacting system $\psi_i (\vec{r})$ via $n (\vec{r}) \ = \ \sum_{i=1}^{N} \ |\psi_i (\vec{r})|^2$.

Now, the expectation value $E[n]$ of the Hamiltonian is minimized with respect to the single particle states under normalization constraint, finally yielding the Kohn--Sham equations~\cite{Kohn1965}:
\begin{multline}
\left\{ \ -\frac{\hbar^2}{2m} \nabla^2 \ + \ v_{ext}(\vec{r}) \ 
+ \ v_{\rm H}(\vec{r}) \ + \ v_{\rm xc}(\vec{r}) \ \right\} \ \psi_i(\vec{r})
\\  = \ \varepsilon_i \ \psi_i(\vec{r}) \ 
\label{eq:KS}
\end{multline}

In the above expression, $v_{\rm H}(\vec{r})$ represents the Hartree potential corresponding to the classical electrostatic interaction within an electron cloud, whereas $v_{ext}(\vec{r})$ stands for the external potential that is determined by the nuclei. Finally, the so--called exchange--correlation potential, $v_{\rm xc}(\vec{r})$, then accounts for all quantum--mechanical many--body effects. The Kohn--Sham equations actually correspond to eigenvalue equations that need to be solved self--consistently, as the solutions also (re--)enter the Hamiltonian via the Hartree and the exchange-correlation potential. The total energy of the electronic Hamiltonian is then finally obtained from the following expression:
\begin{equation}
E \ = \ \sum_{i=1}^N \ \varepsilon_i \ + \ E_{\rm xc} [n] \ 
- \ \int \ v_{\rm xc} (\vec{r}) n(\vec{r})  \ d\vec{r}
- \ V_{\rm H}[n] 
\label{EKS}
\end{equation}

At this point it is interesting to note that neglecting the exchange--correlation functional $E_{\rm xc}[n]$ and its functional derivative $v_{\rm xc}[n]$ would simply result in the Hartree approximation, \textit{i.e.}, the energy of a system without any quantum-mechanical many-body effects. Note furthermore that the eigenvalues of the Kohn--Sham equation $\varepsilon_i$ are typically interpreted as single particle energies, which has proven to be a valid approximation. However, in principle, the physical meaning of the $\varepsilon_i$ is not a priori clear, as they originally correspond the Lagrangian multipliers in the variational approach used to determine the ground state energy of the Hamiltonian. 

Indeed, as can be inferred from Eq.(\ref{EKS}), the exchange--correlation term crucially contributes to the total energy of the system. As already stated above, it contains all quantum-mechanical many-body effects, and only a suitable approximation of this unknown term allows for realistic modelling. 
With the advent of DFT, the exchange--correlation was originally described within the so--called local density approximation (LDA), which is based on the electron density and the corresponding exchange--correlation energy of the homogeneous electron gas~\cite{Kohn1965}. This means that within a LDA calculation only the local electron density is taken into account, which makes LDA a local functional:
\begin{equation}
E_{\rm xc}^{\rm LDA} [n] \ = \ \int d^3\vec{r} \ n(\vec{r}) 
\  \varepsilon_{\rm xc}^{\rm LDA}(n(\vec{r})) 
\label{LDA}
\end{equation}

The standard approach in material science is based on the generalized gradient approximation (GGA)~\cite{Langreth1983,Becke1988,Perdew1992}, for which the Perdew-Burke-Ernzerhof (PBE) functional is the by far most frequently used implementation~\cite{Perdew1996}. The GGA formalism describes the exchange--correlation as functional of the local electron density and the gradient of the latter one, meaning that $\varepsilon_{\rm xc}^{\rm LDA}$ is replaced by $\varepsilon_{\rm xc}^{\rm GGA}(n(\vec{r}), | \nabla n(\vec{r})|)$.
This type of functional is called semi--local, as the consideration of the density gradient indirectly accounts for the nearby electron density. While GGA calculations certainly have yielded satisfactory results in many applications, there remain shortcomings of this approximations, which may become rather severe, depending on the system. In particular the treatment of systems with localized electrons, such as the {\it{d}}--electrons in transition metals, is prone to errors in GGA based calculations since this functional tends to unrealistically smear out the electron density~\cite{Lundberg2005}. This is a consequence of the so--called self--interaction error that can be shown to result in an artificial delocalization of electrons for semi--local functionals. Going along with this fact, GGA calculations are not able to correctly reproduce the band gaps of semi--conductors and insulators. 

However, these shortcomings can be corrected in a simple and efficient way by applying the GGA+U approach. Indeed, this means that the more localized character of electrons can simply be accounted for by a Hubbard--model like correction~\cite{Anisimov1991}. Here, often the rotationally invariant form as introduced by Dudarev is applied~\cite{Dudarev1998}. While GGA+U in principle is simple and efficient, it relies on a parameter (the U parameter) that has to be supplied to the calculation either as an empirical value or can be obtained via a linear response calculation for the given system. In practice, often empirical U parameters are used, which, however, in a way makes the calculations lose their predictive power.
For Li--ion batteries, benchmark studies have been conducted that were able to predict a reliable set of U parameters in particular for the treatment of oxide materials~\cite{Jain2011}.
Still, the search for improved descriptions of the exchange--correlation has been an ongoing task and methods beyond GGA and GGA+U have been developed. The conceptually next step is to include not only the gradient of the electron density but also higher order terms, such as the Laplacian or the kinetic energy density, for the description of the exchange correlation~\cite{Neumann1997,Perdew1999,Tao2003}. Again, there exist different implementations of these so--called meta--GGA functionals. Here, the rather recently developed SCAN functional seems to be a highly promising variant which describes many situations correctly at the expense of much lower computational cost than calculations for instance based on hybrid functionals~\cite{Sun2015a}. The latter ones are constructed by including a specific amount of exact exchange, meaning that a certain portion of the wave--function based Hartree--Fock exchange is considered, making this type of calculations significantly more expensive. Again there exist different implementations such as HSE, B3LYP or PBE0, which mostly differ by the amount of exact--exchange that is considered~\cite{Heyd2003,Kim1994,Stephens1994,Adamo1999}.

Finally, it has to be pointed out that most functionals are error--prone with respect to systems with strong van der Waals interactions, which for instance is crucial for the adsorption of molecules or the interlayer interaction in graphite. Hence, the correct modelling of van der Waals interactions may indeed also be of importance for certain systems in battery research. 
There exist various approaches to include dispersion forces, one of them being the Grimme D3 method~\cite{Grimme2010}. It provides a simple correction scheme that is computationally essentially free of cost. Indeed, the D3 correction simply uses a set of parameters that has been determined for particular molecules to describe the van der Waals interaction. Despite the conceptually simple approach, there are many cases, where this scheme has proven to be quite accurate~\cite{Mahlberg2019}. For surfaces, additional improvement has been obtained when in combination with the dispersion correction the modified RPBE functional is used, which is an adapted version of the PBE functional~\cite{Hammer1999,Forster-Tonigold2014}.
A different way to treat van der Waals interactions is by taking the non--local nature of van der Waals forces into account, applying a non--local functional. While this approach has also been successfully applied for battery materials, it comes at somewhat increased computational cost~\cite{Dion2004}. 

A further functional alternative that has to be mentioned in this context is the BEEF--vdW functional (Bayesian Error Estimation Functional). It is a semi--local functional that includes a non--local correction term and was designed as a general purpose exchange--correlation functional. The BEEF functional is obtained from a machine learning inspired approach and has to be understood as a compromise to best describe the different kinds of physical and chemical interactions, particularly suited for catalysis and surface science~\cite{Wellendorff2012}.
Moreover, the BEEF functional allows for an uncertainty quantification at minimal computational cost by determining the total energy for an ensemble of functionals (with a defined distribution of different model parameters)~\cite{Wellendorff2012}.

\begin{figure}
\includegraphics[angle=0,width=0.75\columnwidth]{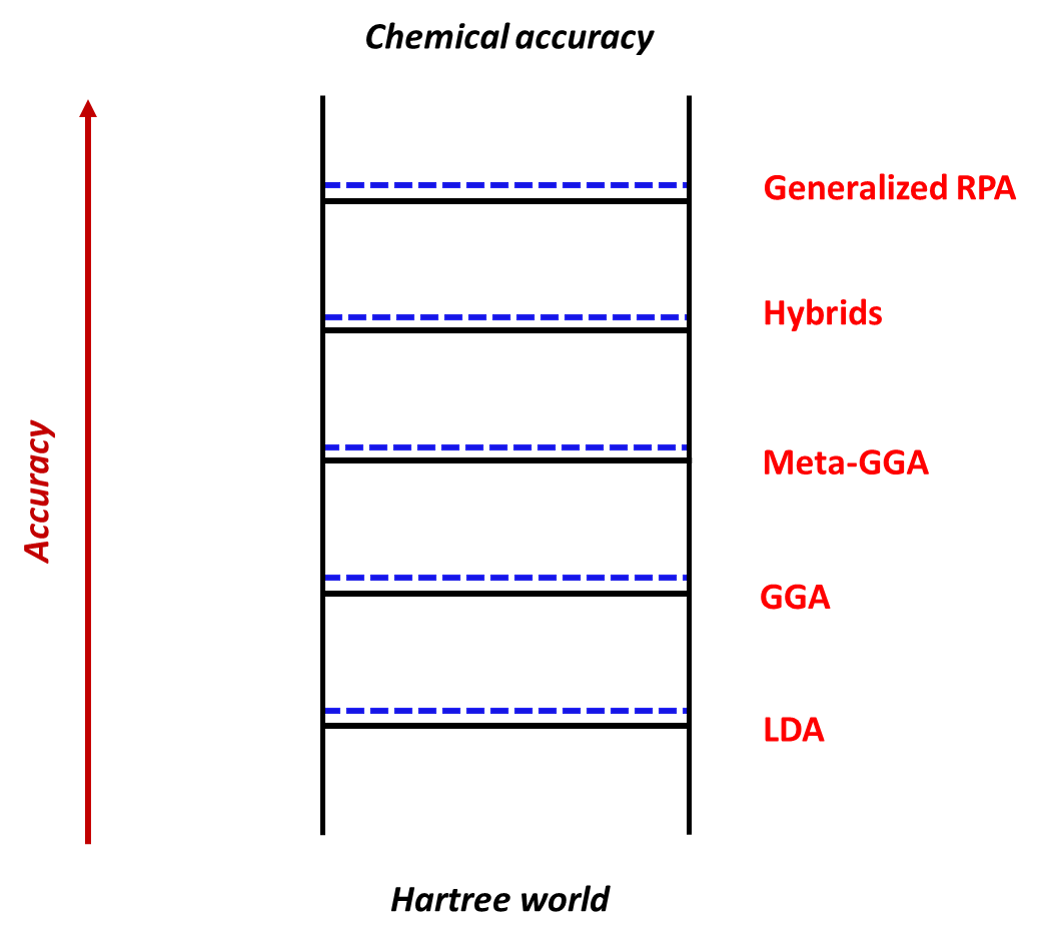}
\caption{Jacobs ladder schematically picturing the evolution of exchange--correlation functionals to reach an increasing accuracy~\cite{Perdew2001}. The blue dashed lines indicate van der Waals type corrections for the respective levels.}
\label{fig:Jacobs-ladder}
\end{figure}

Finally, the search for an improved description of the exchange--correlation functional is still an ongoing task, ideally striving for achieving chemical accuracy, which is typically defined as $\approx$ 0.04 eV per atom (1kcal/mol).
When referring to the development of more accurate functionals that are able to describe different types of problems, often the picture of the Jacob's ladder as introduced by Perdew is evoked~\cite{Perdew2001}.
This ladder simply symbolizes that a more accurate description of the exchange--correlation functional, meaning amongst others the fulfillment of more mathematical constraints, is desirable and also accessible with increasing computer power (see Fig.~\ref{fig:Jacobs-ladder}). 
However, it has to be pointed out that there is no straight-forward and systematic way of improvement, which is maybe a bit misleadingly suggested by this picture of a ladder. In fact, for a particular problem it may be possible that a simple description in terms of the most widely used PBE functional is more successful than a highly expensive hybrid functional calculation. This is for instance true for metals, where the inclusion of exact exchange may lead to artifacts in the density of states~\cite{Paier2007}. 

\begin{figure}
\includegraphics[angle=0,width=0.95\columnwidth]{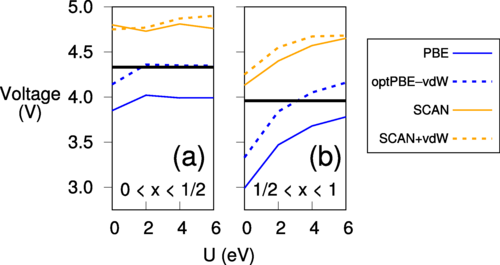}
\caption{Insertion voltages for Li$_x$CoO$_2$ as obtained by applying different exchange--correlation functionals as function of the U parameter. Reprinted with permission from Isaacs et al.~\cite{Isaacs2020}. Copyright (2020) American Physical Society.}
\label{fig:comp-functionals}
\end{figure}

To exemplify the quantitative impact of different exchange--correlation functionals, the calculated insertion voltages in Li$_x$CoO$_2$ is depicted for selected functionals and as a function of the U paramer (see Fig.~\ref{fig:comp-functionals}). While the results clearly show quantitative differences, it has to be pointed out that qualitative trends are often the same.  
Hence, it should be emphasized that the best suited functional indeed depends on the exact problem and therefore has to be chosen with care, both with respect to a suitable description of the system under investigation and reasonable computational cost.

In practice, any DFT approach starts with the structural optimization of the investigated compound. After determining the minimum energy configuration, the conceptually next step is to determine the properties one is interested in. Apart from energetic stability with respect to competing phases, the electronic band structure, optical, mechanical, vibrational and many more material properties can be obtained. This means that DFT on the one hand has a high predictive power and on the other is a versatile tool to interpret experimental results by a full analysis of the structure--property relationships of a given compound.

\subsubsection{Electronic Structure}

As discussed above, the eigenvalues of the Kohn--Sham equations correspond to the eigenenergies of the respective single electron wave functions. While it is not obvious that these eigenenergies have a physical meaning, in practice these single electron states are used to determine the band structure and the electronic density of states (DOS) of compounds under investigation. The such obtained band structure and DOS are usually successfully interpreted as physical quantities and may be used to gain insight into bonding character and electronic stability of the investigated material. This can be justified by the fact that in many cases the one-particle energies determined by more advanced schemes such as the GW approximation only differ by an approximately constant shift from the Kohn-Sham eigenvalues~\cite{Rohlfing95} so that the shape of the band structure is hardly affected except for the band gap.

\begin{figure}
\includegraphics[angle=0,width=0.95\columnwidth]{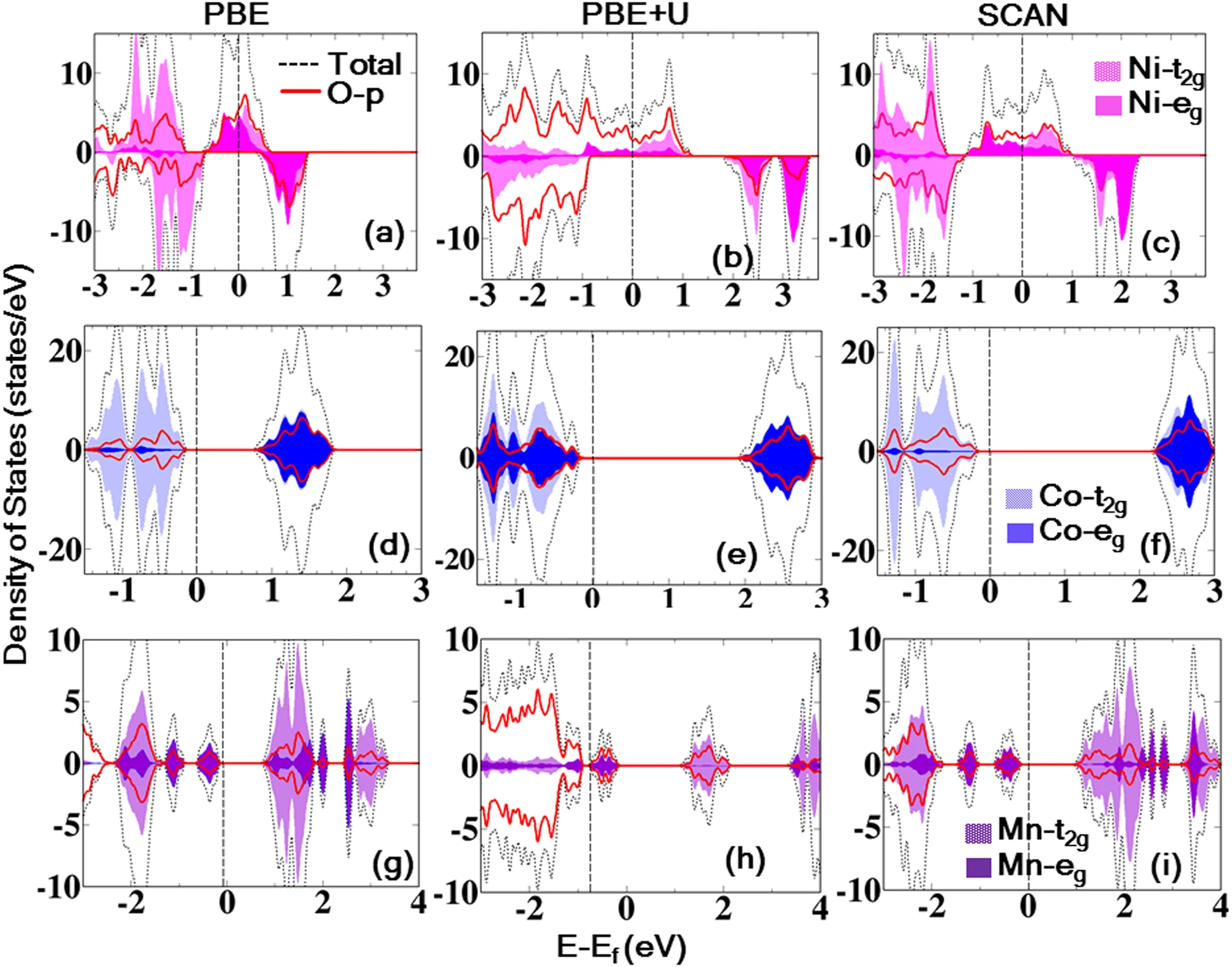}
\caption{Density of states for LiTMO$_2$ (TM = Ni, Co, Mn), as obtained from DFT calculation applying the PBE, PBE+U, or SCAN functionals. From Chakraborty et al.~\cite{Chakraborty2018}. This article is licensed under a Creative Commons Attribution 4.0 International License.} 
\label{fig:Chakraborty}
\end{figure}

A projection of the eigenstates on atom--centered orbitals furthermore allows a detailed analysis of the bonding situation with respect to \textit{s--, p--, d--} and \textit{f--} type features. In Fig.~\ref{fig:Chakraborty} the DOS of different Li--intercalated layered oxides and the projections on oxygen 2$p$-- and TM 3$d$-- states is depicted, emphasizing differences originating from the selected TM species, but also illustrating once more the impact of the applied exchange--correlation functional. 
In addition, advanced analysis schemes to distinguish \textit{e.g.}, bonding from non--bonding and anti--bonding contributions do exist, such as the crystal overlap orbital population (COOP) and electron localization function (ELF)\cite{Dronskowski1993,Becke1990}. These quantities have already been briefly introduced during the discussion of anionic redox processes (see Figs.~\ref{fig:COOP} and~\ref{fig:ELF}).

Moreover, the charge density, especially when visualized as a charge density difference plot, can also provide viable information on the bonding situation in a given material. With respect to redox concepts it is, on the other hand, often desirable to assign a certain charge to a given atom. For this purpose different charge partitioning schemes are available, such as Mulliken charges, Bader charges or the density derived electrostatic and chemical (DDEC) approach~\cite{Mulliken1955,Bader1994,Manz2016}. These methods use certain criteria to assign the charge distribution between the atoms to one atom or the other. While these approaches usually yield the same trends, the absolute numbers may differ strongly. In particular, these methods will typically not yield integer numbers such that there is no one to one correspondence between the DFT computed charge and the assigned oxidation state. Nevertheless, these charge partitioning schemes are extremely useful to track down changes in the charge distribution -- \textit{e.g.}, under de--/lithiation of a compound.

\subsubsection{Lattice Dynamics}

Determining phase diagrams of solids with respect to temperature in principle means that the free energy has to be evaluated. However, in most cases temperature dependent contributions are neglected and hence phase diagrams often are determined without taking vibrational degrees of freedom into account. This is mostly due to the fact that these contributions are rather small and that accessing vibrational properties results in additional computational costs. On the other hand, for a detailed investigation of a compound the vibrational properties may be of great interest, as they contain viable information and can, moreover, be an extremely sensitive measure for the accuracy of the model description. 
Experimentally, Raman spectroscopy and in particular \textit{operando} Raman studies on electrodes, have become an important lab scale tool for the analysis of battery materials. The fact that Raman spectroscopy is a local probe makes it a complementary technique to standard characterization tools such as X--ray diffraction. In principle, a Raman measurement simply gives access to the atomic vibrations or phonons at the $\Gamma$--point, which are directly accessible by means of DFT calculations and can therefore be used to characterize the occurring phases during charge/discharge.
Computationally, this means that the phonon modes at $\Gamma$--point have to be determined, which then can be classified as Raman active depending on their underlying symmetry. By means of perturbation theory, it is then also possible to obtain the Raman intensity of a given mode. In general, phonons are calculated within the harmonic approximation, assuming that the associated vibrations correspond to small displacements of the atoms out of their equilibrium. For this case, the energy landscape can be approximated by a harmonic potential. 

From a generalized point of view, the theoretical treatment starts from the following Hamiltonian: 
\begin{equation}
 H \ = \ T_{nucl} \ + \ V_{BO}(\{\vec{R}_I\})
\label{eq:BO-phonon}
\end{equation}
This formulation corresponds to describing the lattice vibrations as atomic motions on the Born--Oppenheimer potential energy surface.
Expressing the displacement out of the equilibrium position $\vec{R}_I^0$ by a displacement vector $\vec{u}_I(\vec{R}_I^0)$, the potential energy can be expanded in a Taylor series with respect to the equilibrium, thus yielding:
\begin{multline}
  V_{BO}(\{\vec{R}_I\}) \ =  \ E_{el}(\{\vec{R}_I^0\}) \ + \\ \ \frac{1}{2}\sum_{I,J} \sum_{\mu, \nu}\frac{\partial^2 E_{el}(\{\vec{R}_I^0\})}{\partial {u_I^{\mu}} \partial{u_J^{\nu}}} {u_I^{\mu}}{u_J^{\nu}} \ + \ ...
  \label{eq:Taylor}
\end{multline}
As the Taylor expansion is constructed around the equilibrium positions, there are no forces acting on the atoms and therefore the terms containing the first derivatives have to disappear. Treating the kinetic energy as classical quantity, one finally obtains an expression for the system Hamiltonian that corresponds to a set of coupled harmonic oscillators:
\begin{multline}
H = \sum_{I} \frac{M_I}{2}\dot{\vec{u}}_I^2 \ + \ E_{el}(\{\vec{R}_I^0\}) \\ \ +  \ \frac{1}{2}\sum_{I,J} \sum_{\mu, \nu}  \Phi_{\mu\nu}(\vec{R}_I^0,\vec{R}_J^0) u_I^{\mu} u_J^{\nu},
\end{multline}
with the harmonic force constants
\begin{equation}
  \Phi_{\mu\nu}(\vec{R}_I^0,\vec{R}_J^0) = \frac{\partial^2 E_{el}(\{\vec{R}_I^0\})}{\partial u_I^{\mu}\partial u_J^{\nu}},
\end{equation}
  as obtained from the Taylor expansion in Eq.~(\ref{eq:Taylor}). The corresponding equation of motion can be solved by imposing periodic boundary conditions and using a plane wave ansatz for the displacements, which finally reduces to the eigenvalue equation

\begin{equation}
 \sum_{J,\nu} D_{IJ}^{\mu\nu} (\vec{q}) \varepsilon_J^{\nu} (\vec{q}) \ = \ \omega^2\, \varepsilon_J^{\mu}(\vec{q}),
\end{equation}

\noindent with the dynamical matrix $D_{IJ}^{\mu\nu}$:  
\begin{equation}
 D_{IJ}^{\mu\nu} \ = \ \frac{1}{\sqrt{M_{I}M_{J}}} \sum_n \Phi_{IJ}^{\mu\nu}(\vec{R}_n)e^{i\vec{q}\cdot\vec{R}_{nI}}
\end{equation}

By solving this eigenvalue problem for a distinct wave vector $\vec{q}$, the corresponding phonon frequencies can directly be obtained. Hence, the main task is indeed to determine the dynamical matrix, which can be achieved by different approaches. First, there are density functional perturbation theory (DFPT)~\cite{Baroni2001} calculations and second there is the finite displacement approach, which is often referred to as direct method~\cite{Frank1995,Parlinski1997}. The first approach has the advantage that it can be performed on the unit cell of the system, however, the DFPT calculation then has to be conducted for each desired $\vec{q}$--point, separately. The direct method typically uses finite displacements of symmetry--non--equivalent atoms to determine the force constants. Here, a supercell has to be used, to make sure that the dynamics of the system is captured correctly, since otherwise spurious self--interactions may occur. However, usually a supercell size below 8--10 {\AA} has proven to be sufficient. 

\begin{figure}
\includegraphics[angle=0,width=0.95\columnwidth]{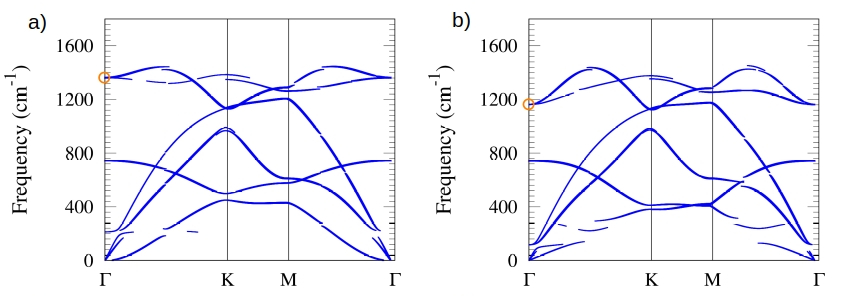}
\caption{Phonon dispersion curves for LiC$_6$ and NaC$_6$, projected to the reciprocal lattice of graphite. The yellow circle at $\Gamma$--point indicates the position of the Raman active G--band~\cite{AnjiReddy2018,Euchner2020a}.}
\label{fig:phonon}
\end{figure}

As the vibrational frequencies depend on the specific wave vector for which the dynamical matrix is solved, one typically determines solutions along a certain path in reciprocal space. This yields the phonon dispersion curves, which are conceptually closely related to the electronic band structure. In Fig.~\ref{fig:phonon}, the dispersion curves for LiC$_6$ and NaC$_6$ are depicted~\cite{AnjiReddy2018,Euchner2020a}, clearly showing that differences between Li-- and Na-- intercalation compounds are imprinted in their vibrational spectra. The yellow circle indicates the so--called G--band -- a characteristic Raman signature of graphitic compounds -- which is observed to shift towards lower frequencies under Li/Na--insertion. This has lead to identifying the shifting G--band position as signature of the intercalation process~\cite{AnjiReddy2018,Euchner2020a}. On the other hand, instead of investigating distinct directions, a sampling of reciprocal space may be of interest. This gives access to the vibrational density of states, which actually determines quantities such as for instance the specific heat.

An important side note with respect to the computational treatment of lattice vibrations is, moreover, that in polar materials a non--analytical correction has to be considered to correctly account for the LO/TO splitting at the zone center. For this purpose, the additional term
\begin{equation}
  D_{IJ}^{\mu\nu,NA} \ = \ \frac{1}{\sqrt{M_{I}M_{J}}} \ \frac{4\pi e^2}{N\Omega} \ \frac{(\vec{q}\cdot Z_{\mu}^*)_{I}(\vec{q}\cdot Z_{\nu}^*)_{J}}{\vec{q}\cdot\boldsymbol{\epsilon}^{\infty}\cdot\vec{q}} \ \\ \ \times \sum_n e^{i\vec{q}\cdot\vec{R}_n},
  \label{eq:nac}
\end{equation}
with $\boldsymbol{\epsilon}^{\infty}$ and $Z_{\mu}^*$ being the dielectric and the Born effective charge tensor, has to be added to the dynamical matrix~\cite{Baroni2001, Gonze1997, Cochran1962}. However, the quantities contained in Eq.~(\ref{eq:nac}) can in principle easily be determined by density functional perturbation theory~\cite{Gonze1997,Giannozzi1991}.

Before concluding this paragraph, it has once more to be pointed out that, apart from being a sensitive probe for gaining insight into structure and dynamics of a material, phonons are contributing to the free energy of a compound. Indeed, with respect to free energy calculations, the theoretical treatment rarely goes beyond the limits of the just introduced harmonic approximation, meaning that anharmonic effects are typically neglected. Nevertheless, anharmonicities -- \textit{i.e.}, deviations of the potential landscape from a quadratic form -- may become important, in particular with increasing temperature.
They are, apart from resulting in frequency shifts and a broadening of the vibrational spectra, also of interest for macroscopic properties, as anharmonicities are for instance responsible for the finite lattice thermal conductivity or the thermal expansion of solids.
In principle, there are different ways of handling the anharmonic contributions to the free energy with respect to temperature and volume. The first, rather obvious and computationally least expensive way is the frequently applied quasiharmonic approximation (QHA). For this approach, harmonic calculations are performed at different cell volumes, thus allowing to assess the quasiharmonic contribution to the free energy F$^{qha}$(V,T). Yet, extensions beyond the QHA exist that are based on the determination and perturbative treatment of higher order force constants, thus giving access to temperature dependent frequency renormalizations. Such frequency renormalizations can be achieved within the self--consistent phonon (SCP) theory or its extension, the improved self--consistent (ISC) theory~\cite{Werthamer1970,Tadano2015a,Oba2019}. In principle, \textit{ab inito} MD may also be applied to determine dispersion curves or the phonon density of states through the use of the Fourier transform of the velocity autocorrelation function~\cite{Forster-Tonigold2013}. This approach implicitly accounts for anharmonic effects, however, means long simulation runs and rather large supercells to obtain suitable resolution in energy and reciprocal space. Finally, there exist advanced methods to determine the anharmonic corrections to the free energy without explicit consideration of higher order force constants, one of them being thermodynamic integration~\cite{Vocadlo2002,Glensk2015, Grabowski2019}.

\subsubsection{Disorder}
\label{sec:SQS}

\begin{figure}
\includegraphics[angle=0,width=0.95\columnwidth]{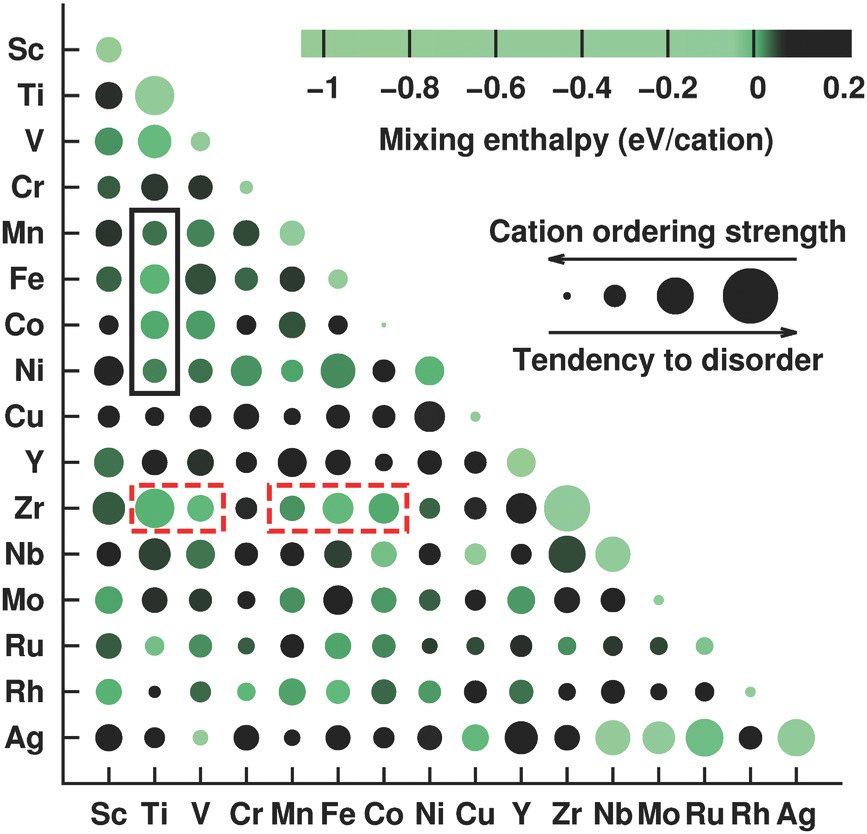}
\caption{Predicted tendency to form a disordered rocksalt compound as determined from comparison between fully ordered and SQS structures of LiA$_{0.5}$B$_{0.5}$O$_2$ stoichiometry. The size of the circle represents the ordering strength (small circles indicate strong ordering tendencies, whereas large ones indicate compounds that are likely to be disordered). The color of the circles quantifies the thermodynamic stability with respect to competing phases (bright colors represent an increased stability).
Used with permission from Urban et al.~\cite{Urban2016d}. Copyright (2016) Wiley--VCH.}
\label{fig:disordering-strength}
\end{figure}

Despite the fact that the computer power has been largely increased, the treatment of disorder still easily exceeds the feasibility of DFT standard calculations, as they rely on relatively small periodically repeated unit cells. A computationally less expensive way of handling disorder in lattice based materials is the use of cluster expansion methods that describe the interaction in solids in terms of adjusted one--particle and truncated many--particle interactions. This in turn enables to project the disorder on large and hence rather realistic simulation cells. On the other hand, developing a cluster expansion for a given compound is also a non--trivial task and may come with the drawback of describing certain situations less accurate than in a DFT calculation. To bridge this gap, a different and rather elegant way to handle disorder was introduced by Zunger~\cite{Zunger1990}. This so--called special quasirandom structure (SQS) approach, allows to create rather small supercells, which aim at fulfilling the mathematical constraints characteristic for a random alloy. In fact, the underlying idea is to alter the occupation of the lattice sites that exhibit disorder until the resulting multisite correlation functions (pairs, triplets etc.) match those in the random limit as close as possible. Hence, the disorder can be mimicked with comparatively small cell sizes by matching these correlations.
Mathematically, this simply means that the corresponding correlation functions $\bar{\Pi}_{k,m}$ have to be calculated. Here $k$ and $m$ define geometric figures, which have $k$ vertices and extend up to the $m^{th}$ nearest neighbour, \textit{i.e.}, single sites, pairs, triplets and so on are considered. Then, similar to the below discussed cluster expansion approach, pseudo spin variables are assigned to the respective atom types~\cite{Jiang2004}.
Finally, the product of these pseudo spin variables for all sites of a figure is calculated, and subsequently the average over equivalent figures is taken, finally yielding the correlation functions $\bar{\Pi}_{k,m}$~\cite{Jiang2004}. 
In the SQS approach a Monte Carlo algorithm is then applied to match the analytical value of the correlation function of a random alloy as close as possible, \textit{i.e.}, $(\bar{\Pi}_{k,m})_{_{SQS}} \cong \langle\bar{\Pi}_{k,m}\rangle_{_R}$~\cite{Jiang2004}. Indeed, this methodology has been shown to work reliably for various types of systems~\cite{Urban2016d, Euchner2015, Klimashin2016,Hahn2019}.
The great advantage of this approach is the fact that it allows a quick construction of representative, moderate sized random structures that can be evaluated by DFT as for instance exploited in a screening study by Urban et al.~\cite{Urban2014}. In this work the energy difference between SQS structures and the most stable ordered arrangements has been used for the assessment of disordering tendencies in Li--TM oxides with LiA$_{0.5}$B$_{0.5}$O$_2$ stoichiometry (see Fig.~\ref{fig:disordering-strength}).
However, while a cluster expansion approach will also be able to detect local ordering, the SQS approach is based on the assumption of a complete random alloy and is hence not suited to treat systems that show short--range ordering. Therefore, if short--range order is to be investigated or \textit{e.g.}, order--disorder phase transitions are of interest, cluster expansion methods are indeed the way to go.

\subsubsection{Defects}

Structural defects are an inherent material property and every realistic material contains a certain amount thereof, such as vacancies, interstitials or anti--site defects, already simply for entropic reasons.
While crystals have to be charge--neutral on a macroscopic scale, these microscopic defects can be positively or negatively charged. To assess the respective stability of charged and uncharged defects, the defect formation energies have to be determined. At this point, it is important to note that periodic calculations are only able to treat charge neutral systems, which means that charged defects have to be compensated within the unit cell in periodic DFT calculations. This can be achieved by introducing a uniform background charge, which in turn results in a certain error that has to be accounted for in the expression for the defect formation energy~\cite{Freysoldt2014}:
\begin{multline}
 E^f[X^q] \ = \  E_{tot}[X^q] \ - \ E_{tot}[bulk] \ - \ \sum_i n_i \mu_i \\  \ 
qE_F \ + \ E_{corr}
\end{multline}
Here, $E_{tot}[bulk]$ and $E_{tot}[X^q]$ correspond to the DFT total energies of the defect free case and a defect containing supercell. The charge of the defect ist denoted by $q$, while $\mu_i$ and $n_i$ are chemical potential and number of species $i$. $E_F$ corresponds to the Fermi level and  is given with respect to the valence band maximum. Finally, the correction term $E_{corr}$ is added, accounting for the error due to the uniform background charge.   

\begin{figure}[t]
\raisebox{+0.4\height}{\includegraphics[angle=0,width=0.3\columnwidth]{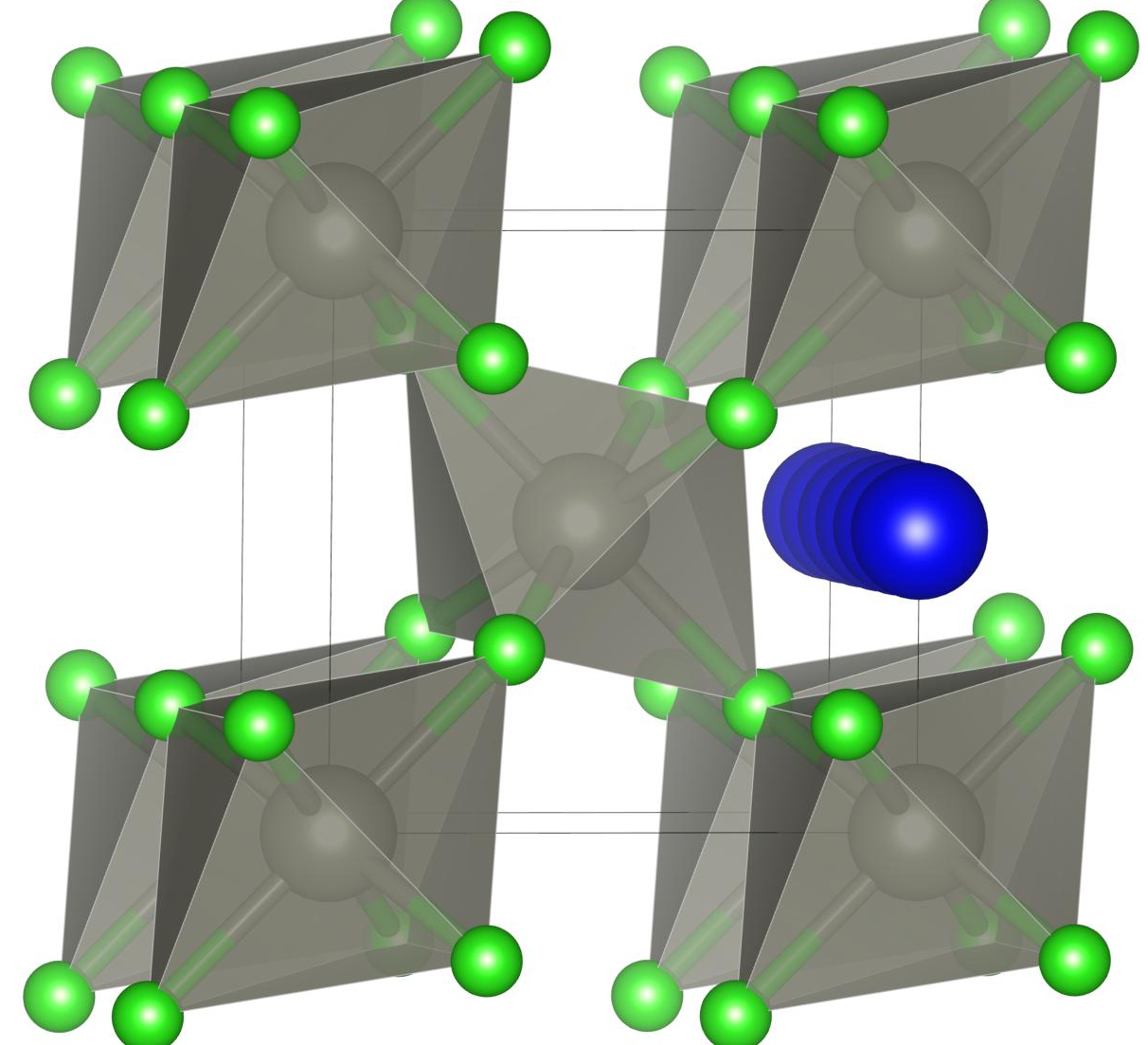}}
\includegraphics[angle=0,width=0.65\columnwidth]{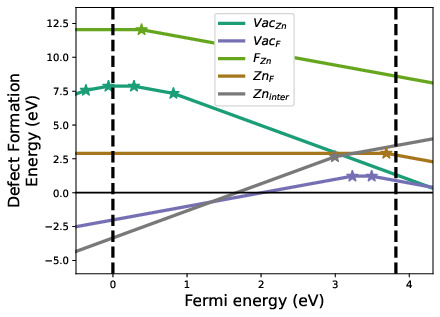}
 \caption{Unit cell of ZnF$_2$, highlighting the available diffusion channels along the c--axis (left). Defect formation energy as a function of the Fermi energy for different types of defects such as vacancies, interstitials and anti--site defects (right). Reprinted with permission from Han et al.~\cite{Han2021}. Copyright (2021) American Chemical Society.} 
 \label{fig:defect}
\end{figure}

While charged defects are an important topic by themselves and extensive reviews on their theoretical treatment exist~\cite{Freysoldt2014}, they may also be crucial when battery materials are investigated. In particular, solid electrolytes need to be bad electronic conductors, meaning that charged defects may become important for the ion conduction in these materials. Such a scenario is discussed for the case of ZnF$_2$, a compound which recently has been suggested as coating material for Zn metal anodes~\cite{Cao2020} and therefore needs to allow for Zn--ion diffusion. Interestingly, the insertion of Zn metal into the empty channels of the ZnF$_2$ structure is energetically unfavorable, such that neutral Zn atoms cannot be expected to enter these channels. However, the calculated defect formation energies show that the formation of Zn$^{2+}$ interstitials is favorable for a broad range of Fermi energies (see Fig.~\ref{fig:defect}~\cite{Han2021}). Hence, under certain electrochemical conditions interstitial Zn$^{2+}$ ions can exist in the empty channels of ZnF$_2$, whereas neutral Zn atoms are thermodynamically largely unstable in the ZnF$_2$ matrix. 
Indeed, this type of considerations may be of particular interest when solid electrolytes are considered. To discuss the possibility of a certain conduction mechanism, it is important to investigate the stability also with respect to the insertion of charged ions. As already stated above, macroscopically a crystal needs to be charge neutral, consequently indicating that the insertion of a positive ion in a solid electrolyte will have to be balanced in the electrode/electrolyte interface region.

\subsubsection{Diffusion}
\label{sec:diffusion}
To enable fast charging of a battery, good kinetics is of particular importance. On an atomistic scale this means that fast migration of the diffusing ions from one lattice site to the other has to be possible. The activation barriers that have to be overcome for such a migration process correspond to free energy differences. With some basic knowledge about the diffusion mechanism -- often intuitive and based on geometrical reasoning -- individual diffusion events can be investigated. To connect atomistic diffusion barriers to a macroscopic diffusivity, transition state theory (TST)~\cite{Haenggi90} is often used. Following a non--rigorous thermodynamic quasi--equilibrium approach, the rate constant of a migration process can be related to the difference in Gibbs free energy between the initial state and the transition state,
as expressed in the frequently applied Eyring equation~\cite{Eyring2013,Evans1935}:
\begin{equation}
 k_{TST} \ = \ \frac{\kappa k_BT}{h} e^{\frac{-\Delta G}{k_BT}}
 \label{eqn:Eyring}
\end{equation}

In a more general theoretical treatment, Vineyard was able to derive the rate constants by using a phase space approach~\cite{Vineyard1957}. Under the assumption of classical dynamics, the rate at which a certain event takes place then finally depends on the probability of reaching the transition state multiplied by the rate of crossing the latter one. The probability of reaching the transition state is obtained from the ratio of the configurational partition functions of initial and transition state, while the rate of crossing is obtained from the average velocity for crossing the transition state, yielding the following expression~\cite{Vineyard1957}:

\begin{equation}
 k_{TST} \ = \ \sqrt{\frac{k_BT}{2\pi}}\frac{\int_S e^{-\Phi/k_BT} dS}{\int_V e^{-\Phi/k_BT} dV}
 \label{eqn:TST}
\end{equation}

Here, $\Phi$ denotes the potential energy as a function of generalized coordinates, while the integrals enclose the phase space volume $V$ surrounding the initial state and the dividing surface $S$, which passes through a saddle point and has to be crossed to reach the final state configurations. 
A frequently used formulation of TST is the harmonic transition state theory (HTST), which additionally assumes a harmonic shape of the potential energy surface in the vicinity of the initial and the transition state. After expanding the potential energy $\Phi$ in Eq.~(\ref{eqn:TST}) into a Taylor series up to second order, the harmonic terms can be expressed with respect to the vibrational normal modes of energy $h\nu$.  These assumptions are justified for $h\nu \gg k_BT$,
finally resulting in the following equation for the rate constant~\cite{Vineyard1957}:

\begin{equation}
 k_{HTST} \ = \ \frac{\prod\limits_{i=1}^{3N}\nu_i}{\prod\limits_{i=1}^{3N-1}\nu_i^{TS}} e^{-\Delta E_A/k_BT} \ = \ \nu^{*} e^{-\Delta E_A/k_BT}
 \label{eqn:HTST}
\end{equation}

$\Delta E_A$ corresponds to the activation energy and refers to the energy difference between initial and transition state.
The term in front of the exponential corresponds to the product of the $3N$ normal mode frequencies of the initial state divided by the $3N-1$ normal mode frequencies at the transition state and is usually expressed as an effective frequency $\nu^{*}$. Note that the transition state corresponds to a saddle point in the potential energy landscape, such that along the reaction coordinate a vibrational mode with imaginary frequency would be observed. Consequently, there are only $3N-1$ normal mode frequencies to be considered for the transition state. In practice, $\nu^{*}$ is typically referred to as pre--exponential factor or attempt frequency. Notably, this frequency term does not correspond to a simple vibrational frequency of the system -- a claim which is often falsely made. In fact, only if the vibrational spectra are essentially unaffected (\textit{i.e.}, they differ only in the additional mode present in the initial state) such an assumption is valid. 
By comparing Eq.~(\ref{eqn:Eyring}) and~(\ref{eqn:HTST}) their similarity is evident and it can indeed be shown that they are essentially equivalent. 

On the macroscopic scale, diffusion processes are typically described by Fick's law:
\begin{equation}
\vec{J} \ = \ -D\nabla c, 
 \label{eq:Fick}
\end{equation}
\noindent which relates the particle flux $\vec{J}$ to the concentration gradient $\nabla c$ via the chemical diffusion coefficient $D$.
In fact, the latter is closely related to the jump diffusion coefficient $D_J$, which can be expressed with respect to the displacements of the diffusing particles throughout time~\cite{vanderVen2001,vanderVen2013}:
\begin{equation}
D \ = \ \theta D_J \ = \ \lim \limits_{n\to\infty} \theta \frac{1}{2d}\frac{\Bigl< \bigl[\frac{1}{N} \sum\limits_{I=1}^N \vec{R}_I(t) \ - \ \vec{R}_I(0)\bigr]^2\Bigr>}{t},
 \label{eq:Einstein1}
\end{equation}
with $d$ the dimensionality of the diffusion process and $\vec{R}_I(t)$ the position of particle $I$ at time $t$.
While $D_J$ corresponds to the jump diffusion coefficient, the additional
prefactor $\theta$ accounts for the fact that strictly speaking the driving force for diffusion is the gradient in chemical potential and not in concentration~\cite{Gomer1990,Uebing1994,vanderVen2001,vanderVen2013}.

Assuming no cross--correlations between displacements of different particles, the time average in Eq.~(\ref{eq:Einstein1}) simplifies to the mean square displacement of the individual particles.
Moreover, in the dilute limit the chemical potential can be assumed to correspond to that one of an ideal solution, thus resulting in $\theta$ being equal to one~\cite{vanderVen2001,vanderVen2013}.
The remaining expression then corresponds to the well--known case of tracer diffusion: 
\begin{equation}
 D^* \ = \ \lim \limits_{n\to\infty} \frac{1}{2d}\frac{\Bigl< \frac{1}{N}\sum\limits_{i=1}^N [\vec{R}_I(t) \ - \ \vec{R}_I(0)]^2\Bigr>}{t}
 \label{eq:Einstein2}
\end{equation}

\begin{figure}[t]
 \center\includegraphics[angle=0,width=0.9\columnwidth]{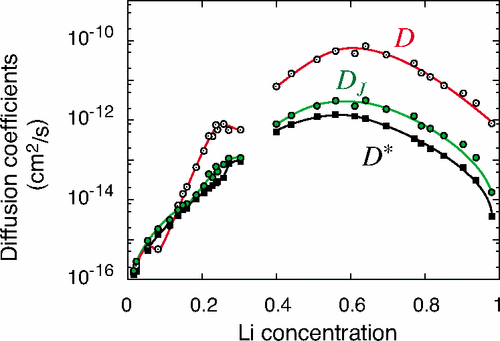}
 \caption{Tracer diffusion (D$^*$), jump diffusion (D$_J$) and chemical diffusion (D) coefficients of layered Li$_x$TiS$_2$ with respect to the Li content, as obtained from kinetic Monte Carlo simulations. Reprinted with permission from van der Ven et al.~\cite{vanderVen2008}. Copyright (2021) American Chemical Society.}
 \label{fig:tracer}
\end{figure}

Hence, in the dilute limit ($\theta\approx1$) and for vanishing cross--correlations $D$ and $D^*$ are identical. To emphasize this point, the concentration dependence of tracer diffusion (D$^*$), jump diffusion (D$_J$) and chemical diffusion (D) coefficient for the case of Li diffusion in Li$_x$TiS$_2$ are depicted in Fig.~\ref{fig:tracer}. While all three show the qualitatively same behaviour, only for vanishing Li--concentration, \textit{i.e.}, in the dilute limit, the same value is approached.
Finally, under the assumption of a random walk on a given lattice, the diffusivity of dilute charge carriers can directly be related to the above discussed rate constants, via~\cite{LeClaire1978,vanderVen2001}:

\begin{equation}
 D \ = \ \alpha^2 g f x_D \nu^{*} e^{-\frac{\Delta E_A}{k_BT}}
\label{eq:dilute}
\end{equation}

Here, $\alpha$ is the hop distance, $g$ the geometry factor, which is related to the underlying lattice, and $f$ a correlation factor (for Markovian motion $f$ is equal to 1). Finally, $x_D$ is the diffusion mediating defect concentration. These prefactors have to be multiplied by the above derived rate constant, consisting of attempt frequency times rate of success.  
Often, the pre--exponential factor $\nu^{*}$ can be assumed to be constant with respect to temperature, which makes the diffusivity obey an Arrhenius law. In practice, the pre--exponential factor is often crudely approximated to be of the order of $~10^{13}$ s$^{-1}$.

While in the case of tracer diffusion the relationship between diffusivity and rate constant is given by Eq.~(\ref{eq:dilute}), for more complicated situations that include cross--correlations, non--dilute concentrations or locally varying diffusion barriers, these quantities can be linked by performing kinetic Monte Carlo (kMC) simulations.    

\begin{figure}[t!]
\includegraphics[angle=0,width=0.98\columnwidth]{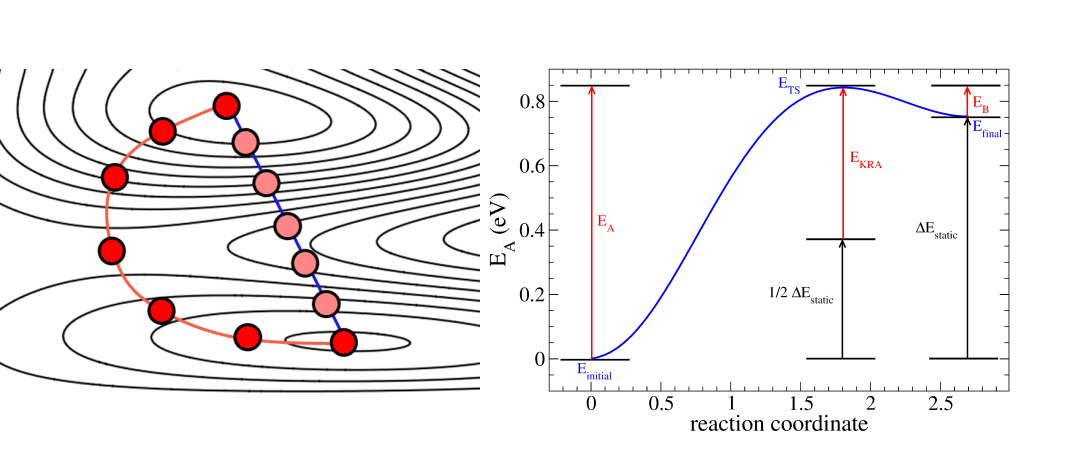}
\caption{
Schematic representation of a NEB chain of so-called images on the corresponding potential energy surface (left). The minimum energy path (MEP), red curved trajectory, is obtained by a constrained optimization of the initial chain that corresponds to a linear interpolation between initial and final state. Kinetically resolved diffusion barrier ($E_{KRA}$) and its construction from initial, final and transition state energy (right). Reprinted from Euchner et al.~\cite{Euchner2020}. Published by The Royal Society of Chemistry.}
\label{fig:E_KRA}
\end{figure}

Now, to gain insight into the diffusion kinetics of a given material via transition state theory, Eq.~(\ref{eqn:HTST}) has to be evaluated. For this purpose, the activation energy $\Delta E_A$, \textit{i.e.} the energy difference between transition and initial state, has to be determined. One way to achieve this in the framework of DFT is the application of the nudged elastic band (NEB) method.
The NEB approach determines the diffusion path with the lowest energy cost and gives access to the corresponding migration barrier. For this purpose, the initial and the final state of a diffusion process must be known. Hence, one must already have some idea about the occurring migration processes.
For the determination of the minimum energy path, the initial and final state are then simply connected by a number of linearly interpolated intermediate configurations, so-called images that are formally connected by virtual springs (see Fig.~\ref{fig:E_KRA}). This initial chain of images is then optimized~\cite{Henkelman2000,Henkelman2000a} involving force projections of both the true
forces and the spring forces, such that ideally the path with the lowest energy cost, connecting initial and final state is obtained, as schematically depicted for the curved line in Fig.~\ref{fig:E_KRA}. However, it needs to be emphasized that the NEB method does not guarantee that the lowest diffusion barrier will be found.

At this stage, it has to be noted that for percolating diffusion pathways in crystalline materials ion migration between non--equivalent sites is rather likely to occur. For such a scenario, it may be of use to separate the barrier in kinetic and static contribution as is achieved by introducing so--called kinetically resolved diffusion barriers
\begin{equation}
 E_{KRA} = E_{TS} - \frac{1}{2}|E_{final} - E_{initial}| \ .
\label{eq:E_KRA}
\end{equation}
Here $E_{TS}$ is the transition state energy, while $E_{final}$ and $E_{initial}$ are the initial and final state energies of the diffusion path. This is exemplified in Fig.~\ref{fig:E_KRA}~\cite{vanderVen2001}, where initial and final state have different site energies, \textit{e.g.}, due to differing local environments. Hence, the barrier actually depends on the direction in which it is to be overcome. The resulting kinetically resolved barriers can be understood as direction independent barriers that describe the kinetics of the diffusion process. Apart from quantifying the kinetic contributions, the kinetically resolved barriers may be used in stochastic approaches to systems that show a distribution of site energies (see \textit{e.g.}, the cluster expansion study of van der Ven et al.~\cite{vanderVen2008}).

A different way to computationally access diffusion properties of a given material by DFT relies on \textit{ab initio} molecular dynamics (AIMD). Here, the most prominent approach is to solve the classical equation of motion for the nuclei, while \textit{ab initio} forces are acting on the latter ones. In short, for a given configuration the forces are determined by solving the electronic Hamiltonian making use of the Hellmann--Feynman theorem. With these forces and an appropriate time step (typically in the order of one fs) the classical equations of motion are solved and the atoms are moved accordingly, as in the case of a classical MD simulation (see below). From the resulting MD trajectories dynamical quantities can be determined. As already discussed above, the diffusivity can be extracted by analyzing the atomic displacements. The corresponding activation barriers can then be derived from the diffusivities obtained at different temperatures, by assuming an Arrhenius type behavior with a diffusion mechanism that is independent of temperature~\cite{Urban2016}.

\begin{figure}[t]
 \center\includegraphics[angle=0,width=0.95\columnwidth]{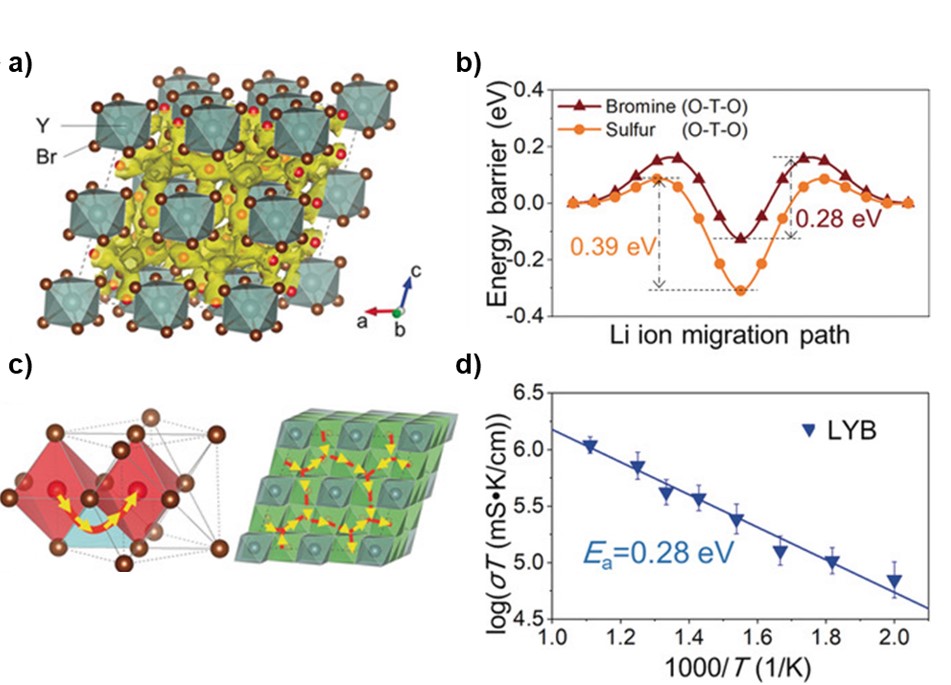}
  \caption{a) Crystal structures and c) Li‐-ion migration pathways in Li$_3$YBr$_6$ (LYB), with the red and blue polyhedrons representing octahedral and tetrahedral interstitial. b) shows the NEB migration barriers for a single Li--ion in the fixed fcc anion lattice of LYB, whereas d) shows an Arrhenius plot of the Li$^+$ diffusivity in LYB as obtained from AIMD simulations.  Adapted with permission from Wang et al.~\cite{Wang2019}. Copyright (2015) Wiley--VCH.}
\label{fig:LYB}
\end{figure}

Before concluding this paragraph, an example that compares the NEB and the AIMD approach for the case of Li diffusion in Li$_3$YBr$_6$ (LYB) will be discussed. In Fig.~\ref{fig:LYB}, the diffusion pathway connecting two octahedral sites through a tetrahedral site (O--T--O) is depicted, with the corresponding NEB barriers shown in panel~b). In panel~d), results of the corresponding AIMD simulations are shown, with the diffusivity as a function of temperature, depicted in logarithmic scale. In this representation it is nicely visible that the diffusivity follows an Arrhenius type behaviour, such that the corresponding diffusion barrier can directly be extracted. The agreement between both approaches is excellent, however, it should be noted that deviations may be observed. This might for instance be due to a slight temperature dependence of the diffusion barriers.

AIMD calculations are computationally rather expensive, however, they can be accelerated by conducting simulations at elevated temperatures. This results in more frequently occurring diffusion events and better statistics, therefore, allowing for shorter simulation times.
As compared to a NEB calculation, AIMD simulations are more computationally demanding and are therefore rather applied if complex mechanism are at play that cannot easily be projected on a NEB trajectory (\textit{e.g.}, concerted motion of several atoms).

\subsection{Classical Molecular Dynamics}
 
If large systems are to be investigated, classical molecular dynamics (MD) is a viable alternative to the above discussed DFT and AIMD approaches. In a classical MD simulation atoms are treated as point-like particles that interact via an effective interaction potential or force field. For these particles, Newtons equation of motion is iteratively solved, meaning that from the knowledge of positions, forces and velocities at a certain time $t_0$ these quantities can be obtained at a later time $t_0 + dt$. The exact way of integrating the equation of motion is determined by the chosen algorithm (\textit{e.g.} the frequently used velocity Verlet algorithm~\cite{Allen2017}).

The numerical integration of the equation of motion allows for the extraction of the respective particle trajectories throughout time~\cite{Allen2017}. Obviously, a classical MD simulation cannot provide information on the electronic structure, but still offers access to many material properties such as phase stability, lattice dynamics or diffusion constants. These quantities can often be determined by evaluating correlation functions, as already discussed for the case of diffusion. This approach allows to explicitly investigate temperature effects such as, for instance, temperature dependent diffusion constants or anharmonicity induced changes in vibrational frequencies. Classical MD simulations can be conducted for much larger system sizes ($10^6$ particles and beyond are easily possible), such that realistic microstructures can be addressed. However, the main factor that decides about the validity of a MD simulation is the quality of the effective interaction potential. If this potential captures the essential features of the interatomic interaction, an accurate simulation will be possible. Here, as in the case of the exchange--correlation functional in DFT, a plethora of different realizations of physically motivated potentials have been developed. As far as metals are concerned, pair potentials have often proven to be sufficient, starting from as simple potential forms as the Lennard--Jones potential, over oscillating pair potentials to more elaborate embedded atom method (EAM) and modified embedded atom method (MEAM) potentials~\cite{Daw1984,Foiles1986,Baskes1992,Baskes1997}.

\begin{figure}[t]
\center\includegraphics[angle=0,width=0.9\columnwidth]{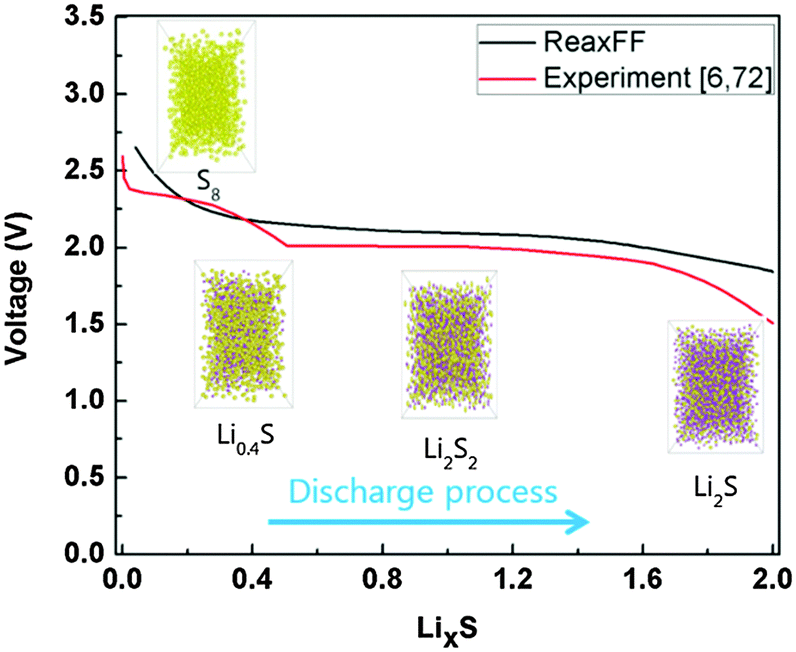}
 \caption{Discharge profile for a Li--S battery as obtained from experiment and from calculations using a reactive force field. Reprinted from Islam et al.~\cite{Islam2015}. Published by The Royal Society of Chemistry.}
 \label{fig:reaxff}
\end{figure}

For the description of covalent bonds, three--body interactions are important and therefore real three--body potentials like the Tersoff potential or pseudo three--body interactions like angular dependent potentials (ADP) have been successfully applied in the past~\cite{Tersoff1989,Schopf2014}. On the other hand, bond breaking and bond creating processes are particularly difficult to describe, thus resulting in the emergence of reactive force fields (ReaxFF), which have also been successfully applied in the field of batteries~\cite{vanDuin2001}. These potentials correspond to bond-order type potentials that are particularly trained to describe bond-breaking and bond-making events as they occur during chemical reactions. An example for a reactive force field study is depicted in Fig.~\ref{fig:reaxff}, where phase evolution and discharge profile for a Li--S cathode have been studied. The MD approach allows here to investigate large system sizes with varying composition and is indeed able to yield excellent agreement with experimental data. Furthermore, there exists a variety of force fields that are especially trained and used for organic molecules, for instance CHARMM, AMBER or GROMOS~\cite{MacKerell1998,Cornell1995,Oostenbrink2004}. All just described potential types have in common that they are based on physical and chemical insights in the materials that shall be investigated, which typically goes along with limited flexibility as far as the description of a broad class of materials is concerned.

Recently, with the advent of machine learning techniques, a new class of potentials has evolved that are not built on physical models but instead rely on complex functions to describe the potential energy landscape, in principle allowing to even reach DFT accuracy~\cite{Lorenz2006, Behler2016}. The basic idea of machine learning (ML) potentials is to represent the input data, \textit{i.e.} the atomic environments in a certain compound, by so--called descriptors. On the basis of these descriptors the ML potential can be created by applying one of the available models to represent the potential energy surface. For the mathematical description of these potential models, artificial neural network potentials (ANN), Gaussian approximated potentials (GAP) or spectral neighbour analysis potentials (SNAP) are popular representatives~\cite{Behler2016}. ANNs are based on neural networks with two or more hidden layers, often using atom centered symmetry functions (ACSFs) as underlying descriptor of the atomic structure. GAPs rely on a Gaussian process kernel, where the kernel may be seen as a similarity measure of atomic environments. As descriptor of these atomic environments, bispectrum components or smoothly overlapping atomic positions (SOAP) are often used to encode the information on the local atomic structure. SNAP potentials, on the other hand, use a linear fitting of the bispectrum components and can in principle be understood as linear version of the GAP model~\cite{Behler2016, Behler2007,Bartok2010,Thompson2015}.

\begin{figure}[t]
 \center\includegraphics[angle=0,width=0.9\columnwidth]{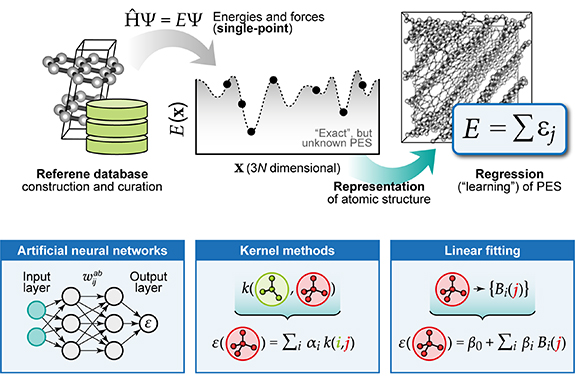}
 \caption{ Schematic picture of the creation of a machine learning potential (top) and characteristic features of three widely used families of machine learning potentials (bottom): (Artificial) neural networks (NN), non--linear kernal function based methods such as Gaussian approximation potentials (GAP) and linearized spectral neighbor analysis potentials (SNAP). Reprinted from Deringer et al.~\cite{Deringer2020}. This article is licensed under a Creative Commons Attribution 4.0 International License.} 
 \label{fig:MLP}
\end{figure}

\begin{figure}[t]
 \includegraphics[angle=0,width=0.98\columnwidth]{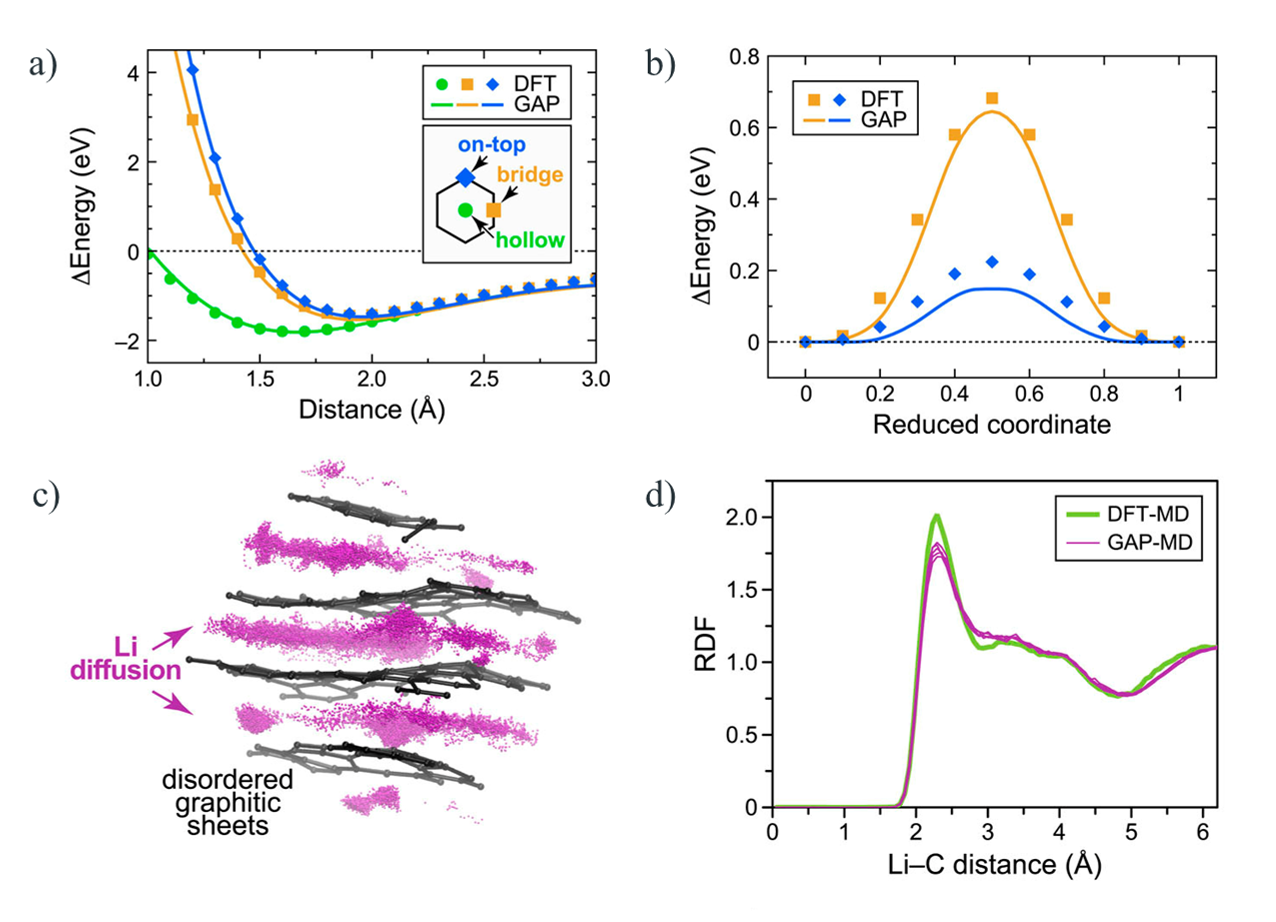}
 \caption{Comparison of DFT and GAP based results for the Li--C system. Panels a) and b) show a comparison for Li adsorption and Li diffusion on graphene. Panel c) depicts the GAP MD trajectory of Li in a graphitic network at a temperature of 1000 K (Li atoms are shown in purple and are repeatedly plotted throughout time, while only the initial framework configuration is shown). Finally, the radial distribution as obtained from AIMD and from machine learning based MD are compared in panel d). Reproduced from Fujikake et al.~\cite{Fujikake2018}, with the permission of AIP Publishing.}
 \label{fig:GAP}
\end{figure}

In Fig.~\ref{fig:MLP}, a schematic representation of the underlying idea and a graphical representation of the just mentioned potential types is depicted.
Clearly, the efficient description of atomistic interactions by ML potentials enables highly accurate investigations on larger length and time scales. This is exemplified in Fig.~\ref{fig:GAP} for the case of the Li--C system. The ML derived GAP potential is able to accurately reproduce DFT data such as adsorption energies and diffusion barriers. Moreover, the distribution of interatomic distances during a GAP MD run also matches almost perfectly with results from AIMD. Hence, this potential allows to study complex geometries -- such as for instance those observed in hard carbon anodes -- and may be used to access the impact of the nano-- and micro--structure with almost DFT accuracy.
The training of such potentials typically needs huge data sets (thousands of DFT calculations), making it a time and resource consuming task~\cite{Lorenz2006}. On the other hand, schemes have evolved for \textit{e.g.} doing an on the fly creation of machine learning potentials during AIMD runs, which thus enables accelerated simulations with high accuracy for smaller system sizes~\cite{Jinnouchi2019} . Generally speaking, MD studies with machine learning potentials are a strongly growing field of research that is also of great interest for the investigation of the complex processes in battery materials. In fact, it seems likely that ML derived potentials will soon dominate newly emerging MD studies, while in many areas of battery research they may even compete with standard DFT approaches. Still, they can not fully replace quantum chemical simulations as ML potentials do not yield information on the underlying electronic structure which is often critical for a deeper understanding of the materials properties.

\subsection{Cluster expansion}

As already stated at several occasions throughout this review, the method of choice to reliably study battery materials from a theoretical point of view is in principle DFT, whenever this is possible. However, there exist many situations where for instance large system sizes or statistical sampling are of interest, which even exceeds the capabilities of MD. For such situations, a coarse graining of the system under investigation may be beneficial, as it can largely reduce the computational cost. In this context, a frequently applied solution are cluster expansion schemes in combination with Monte Carlo methods. For a cluster expansion, the atoms in a system are assumed to occupy a grid of lattice sites with fixed topology, while the species on the respective lattice sites are subject to variations and can be represented by pseudo--spin variables $\sigma_i$. In the case of a binary system (for instance the Li--vacancy arrangement in a layered oxide) $\sigma_i$ may be described by -1 or 1, while for the ternary case (\textit{e.g.} the arrangement of Ni, Mn and Co on the transition metal sublattice of a NMC) $\sigma_i$ could be chosen as -1, 0 or 1 etc. Hence, a certain configuration can simply be described by the corresponding vector $\vec{\sigma} =\{\sigma_1,...,\sigma_n\}$. 

To describe the energetics of such a system, the way how the different species interact with each other needs to be determined. For instance, one might ask if a certain atom prefers to have its own kind or a different atom type as a nearest neighour, next nearest neigbhour and so on. The underlying interactions can then be formulated with respect to structural motives or clusters (points, pairs, triplets, ...), thus allowing to cast the quantum mechanical problem into an effective Hamiltonian. For this purpose, single site basis functions $\theta_n(\sigma_i) $ have to be selected \cite{Sanchez1984,Wolverton1994,Chang2019}. In practice, often Chebyshev polynomials are used for this purpose as they form a complete orthogonal basis. With this basis the so--called cluster functions $\Phi_{\alpha}(\vec{\sigma}) = \prod_i \theta_{n_i}(\sigma_i)$ can be constructed for each cluster motive, finally yielding the following effective Hamiltonian \cite{Sanchez1984,Sanchez2017,vanderVen2020}:
\begin{equation}
 E = \sum_{\alpha} V_{\alpha} \Phi_{\alpha}(\vec{\sigma})
\label{eq:CE}
\end{equation}
The coefficients $V_{\alpha}$ parametrize the effective cluster interactions (ECI) for the different structural motives. Typically, such an expansion is stopped at a triple or quadruple level, where it has to be noted that often not all interactions are important and hence some of them can be disregarded.

For the case of the above mentioned Li--vacancy binary system, the single site basis functions correspond to $\theta_0(\sigma_i) = 1$ and $\theta_1(\sigma_i) = \sigma_i$, such that cluster functions reduce to products of the pseudo spin variables, $\sigma_i, \sigma_i \sigma_j, \sigma_i \sigma_j \sigma_k, ...$, with the indices $i,j,k,...$ running over all lattice sites. The cluster expansion Hamiltonian can then be formulated as:

\begin{multline}
 E = V_0 \ + \ \sum_i V_i\sigma_i \ + \ \sum_{ij} V_{ij}\sigma_i \sigma_j \\ + \ \sum_{ijk} V_{ijk}\sigma_i \sigma_j \sigma_k \ + \ ...,  
\label{eq:CE-binary}
\end{multline}

\noindent with the expansion coefficients $V_i , V_{ij}, V_{ijk}, ...$ representing the ECIs. Note that V$_0$ is a configuration independent term that represents an empty cluster.

The ECIs as a set of parameters are then usually determined by ordinary least square (OLS) fitting -- often amended by a regularization term to prevent overfitting -- to a set of reference data~\cite{Chang2019}. Typically, these reference data are obtained from DFT calculations. In general, the choice of the structural motives that are included is crucial for the quality of the cluster expansion. Indeed, too few motives will yield an inaccurate description of the system, whereas too many may result in overfitting and noise.
A way to find the best compromise is the minimization of the so--called cross validation (CV) score. The CV score allows to select the important motives that are significant for the description of the system and can be understood as an unbiased measure to determine the quality of the cluster expansion with respect to its predictive power towards unknown structures~\cite{vandeWalle2002}. 

\begin{figure}
\includegraphics[angle=0,width=0.98\columnwidth]{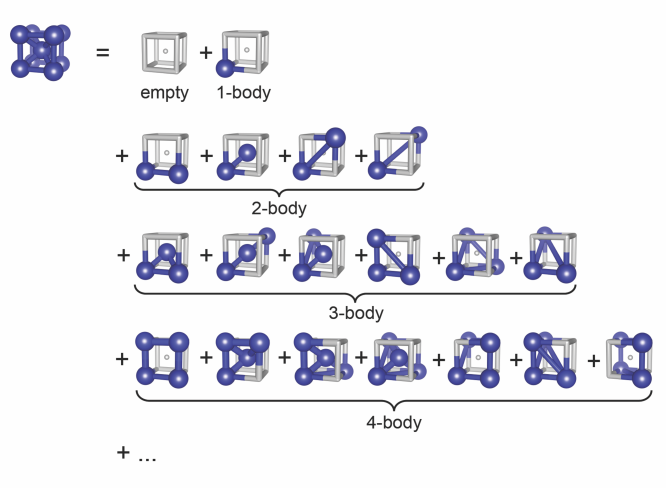}
\caption{3--D representation of different types of structural motives that can be used for constructing a cluster expansion for a bcc lattice. Reprinted from Chang et al.~\cite{Chang2019}. This article is licensed under a Creative Commons Attribution 3.0 International License.\copyright IOP Publishing.} 
\label{fig:CE}
\end{figure}

With such a cluster expansion scheme at hand, the determination of the energy of a given configuration can easily be achieved for large system sizes. In general, thermodynamic properties of a system are determined by the respective microstates through an average thereof. While a cluster expansion enables fast access to the energy of a given configuration, completely accounting for all microstates becomes intractable already for rather small systems. 
Therefore, in a Monte Carlo simulation the phase space has to be sampled, typically applying so--called importance sampling techniques. In practice, frequently the well--known Metropolis algorithm is invoked~\cite{Metropolis1953}, following a general scheme that samples states according to the underlying thermostatistic distribution function, usually corresponding to a canonical or grand canonical ensemble.

In the case of a canonical ensemble, the Metropolis algorithm formally starts with the determination of the energy of a given configuration by evaluating Eq.~(\ref{eq:CE}). As a next step a different configuration is created, \textit{e.g.}, by exchanging two particles. Now, if the energy of this new configuration is lower than that of the previous one it will always be accepted. On the other hand, if the energy is higher than for the previous state, the new configuration is only accepted by a certain probability.
For the Metropolis algorithm, this probability is then given by $e^{-(E_{new}-E_{old})/k_BT}$, such that the new configuration is accepted if this value is smaller than a random number in the range (0,1]. In this way, a distribution of states corresponding to the statistical distribution of states of a canonical ensemble is created (\textit{i.e.}, a Boltzmann distribution). By including the chemical potential in the acceptance probability, this approach can easily be adapted for the grand canonical ensemble as well. While the Metropolis algorithm certainly is the most famous MC algorithm, it has to be noted that for a given problem better suited (faster) algorithms may exist. In particular, at low temperature the Metropolis algorithm is characterized by a high rejection rate, such that different algorithm may result in a considerable speed--up~\cite{Kratzer2009}.  

In practice, cluster expansion based MC simulations can then be applied to determine structural peculiarities, such as ordered superstructures or local short--range ordering~\cite{Sanchez1981,Hinuma2008,Wu2016}, making this approach particularly valuable for the investigation of certain battery materials. The question for the most favorable arrangement of alkali metal ions and vacancies in a layered oxide cathode at a given state of charge may for instance be tackled with such a setup.
Moreover, the temperature dependent stability of different configurations can be investigated, thus among others enabling the determination of concentration--temperature phase diagrams~\cite{Wu2016,Chang2019}. 

\begin{figure}[t]
\center\includegraphics[angle=0,width=0.98\columnwidth]{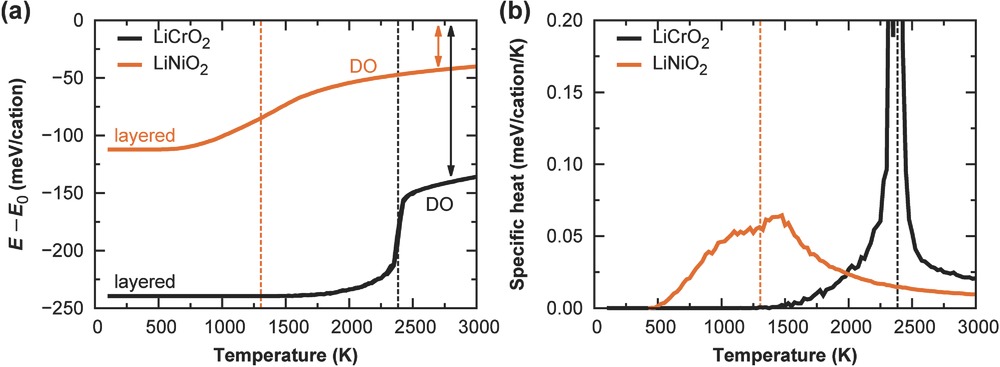}
\caption{a) Energy and b) specific heat of LiCrO$_2$ (orange) and LiNiO$_2$ (black) as obtained from cluster expansion based MC simulations, with the dashed vertical lines yielding the phase transition temperature from the layered to the DRS phase. The arrows indicate the impact of short-‐range order as compared to a fully random cation distribution. Used with permission from Urban et al.~\cite{Urban2016d}. Copyright (2016) Wiley--VCH.}
\label{fig:DRS-kMC}
\end{figure}

As already discussed, MC simulations allow for a correct thermodynamic sampling and hence are able to give access to thermodynamic properties. While the evaluation of Eq.~(\ref{eq:CE}) yields the internal energy of the system, the configurational entropy can be obtained by determining the specific heat and integrating it with respect to temperature. This then allows for the calculation of the corresponding free energy. With the applied thermodynamic sampling, an investigation of phase transitions with respect to temperature becomes directly possible, as exemplified in Fig.~\ref{fig:DRS-kMC}. There, the order/disorder transition of two different layered oxides to the corresponding disordered rocksalt (DRS) oxides -- a recently intensively investigated class of promising cathode materials -- is depicted. For the selected compounds -- LiNiO$_2$ and LiCrO$_2$ -- the cluster expansion based MC simulations nicely show the signature of a first and a second order phase transition, clearly visible when plotting energy and specific heat vs. temperature~\cite{Urban2016d}. 

At this stage, it has to be pointed out that vibrational entropy is usually not considered in cluster expansion based MC simulations, as the relative impact for phases with the same stoichiometry can often be assumed to be rather small~\cite{Urban2016d,vandeWalle2002}. In fact, the configurational entropy differences for order/disorder transitions of a binary alloys are less than $k_\mathrm{B}\ln 2$ per atom ($\approx$ 0.7$k_B$/atom), while typical values for the vibrational entropy differences are of the order of $\approx$ 0.2 $k_B$ per atom~\cite{vandeWalle2002a}. Of course, this nevertheless means that there are cases for which the vibrational contribution indeed becomes important.
Furthermore, MC studies typically also do not consider electronic contributions to the configurational entropy, which may originate from localized electrons. While this is usually justified, there exist cases where this contribution can be crucial for the determination of the phase stability, as \textit{e.g.}, in the case of the Li$_x$FePO$_4$ phase diagram~\cite{Zhou2006}.

\subsection{Kinetic Monte Carlo}

While the above discussed MC methods correspond to a coarse graining of the system of interest and are able to provide information on thermodynamic properties, the underlying kinetics, \textit{i.e.} its time evolution, is not considered. However, as outlined in detail before, often the kinetics of a electrode material is of great interest and coarse grained approaches to the diffusion properties -- in space and time -- are desirable. 

\begin{figure}[t]
\includegraphics[angle=0,width=0.85\columnwidth]{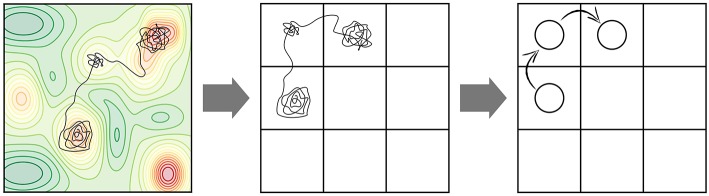}
\caption{a) Potential energy surface (PES) with different local minima and a MD trajectory on this surface. b) Representation of the MD trajectory on a suited lattice. c) Transformation of the MD trajectory in a series of discrete jumps between local minima of the PES. Reprinted from Andersen et al~\cite{Andersen2019}. \copyright 2019 Frontiers Media SA. All Rights Reserved.}
\label{fig:KMC-MD}
\end{figure}

For systems that exhibit a dynamics that is too slow to be captured by AIMD or even classical MD -- meaning that a huge number of time steps would be necessary to capture the events of interest and to obtain a sufficiently accurate statistics -- such coarse graining schemes have to be applied. 
In fact, for this purposes the so--called kinetic Monte Carlo method has been developed. Considering systems with slow dynamics actually means that one is dealing with what is called rare event dynamics. In practice, such systems are assumed to typically oscillate for a long time span around a certain configuration until finally such a rare event (\textit{e.g.}, a diffusion process) takes place. This can then in principle be understood as a separation of time scales of equilibrium oscillations and actual diffusion event (see Fig.~\ref{fig:KMC-MD})~\cite{Reuter2011}.
For a diffusing atom this in turn means that it is oscillating for some time around a local minimum of the potential energy (or Born--Oppenheimer) surface, before moving to an adjacent one. 

A prerequisite for a possible mapping on a kMC simulation is that the investigated diffusion process is a stochastic process with no correlation between successive events. Formally, this means that the investigated process is of Markovian type, such that the system has no memory on how it arrived in a certain state. This is typically true for diffusion in solids, where a diffusing atom usually vibrates around its equilibrium position until it finally jumps to a neighbouring energy minimum. For this situation the dynamics of the system is contained in the corresponding rate constants. These rate constants are then the kinetic parameters that determine the kMC simulation and can be obtained from DFT calculations. In particular, harmonic transition state theory is frequently applied for the determination of rate constants, as discussed earlier. To describe the dynamics of a certain system, all processes for leaving (accessing) a given configuration as well as the corresponding rates $k_{ij}$ need to be determined. The rate $k_{ij}$ then corresponds to the probability per unit time that the system moves from a state $i$ to state $j$. Now, a stochastic description of the kinetics of the whole system in terms of the time evolution of the probabilities is possible and results in the so--called Master equation~\cite{Voter2007,Reuter2011,Andersen2019}:
\begin{equation}
\frac{dP_i}{dt} = -\sum\limits_{i \neq j} k_{ij} P_i(t) + \sum\limits_{i \neq j} k_{ji} P_j(t)  
\end{equation}
Here, the change in probability $P_i$ of finding a certain state $i$ is determined by the probabilities of leaving that state towards a new configuration $j$, as well as by the probabilities that a state $j$ ends up in configuration $i$. Due to the typically large number of states an analytic solution of the Master equation is usually not possible, however, kMC provides an efficient stochastic approach to quantify the kinetics of the system.

The above equation makes it obvious that the kinetics is governed by the rate constants of the respective processes. As long as these rate constants are accurately determined and the processes are indeed not correlated, diffusion constants that are determined from kMC simulations in principle will yield the same result as a much more demanding MD simulation.
In a kMC simulation, the trajectory of a particle then simply consists of a series of discrete hops from one local minimum to the next (see Fig.~\ref{fig:KMC-MD}). The random selection of a given hop and the time span between the hops is governed by the probabilities which have to obey the Master equation 
~\cite{Voter2007,Reuter2011,Andersen2019}. Moreover, the detailed balance criterion is imposed to ensure that the system is in thermodynamic equilibrium~\cite{Fichthorn1991}.

With the system not memorizing its configuration, the probability of leaving a state in a certain time interval is the same as in any previous time interval. This results in the probability for the system having not yet escaped from a given state corresponding to an exponential decay~\cite{Voter2007} -- the survival probability:
\begin{equation}
p_{s}(t) = e^{-k_{tot} t} 
\label{eq:survival}
\end{equation}

With this expression, the probability $p(t)$ of a hop occuring at a certain time can easily be derived from the time derivative of $1-p_{s}(t)$, actually corresponding to a Poisson process with~\cite{Fichthorn1991,Voter2007,Reuter2011}:
\begin{equation}
p(t) = k_{tot} e^{-k_{tot} t}
\label{eq:escape}
\end{equation}

\begin{figure}[t]
\includegraphics[angle=0,width=0.775\columnwidth]{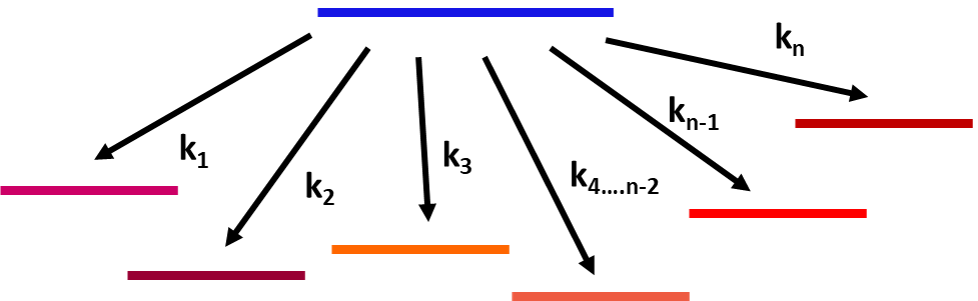}\vspace{0.5cm}
\includegraphics[angle=0,width=0.855\columnwidth]{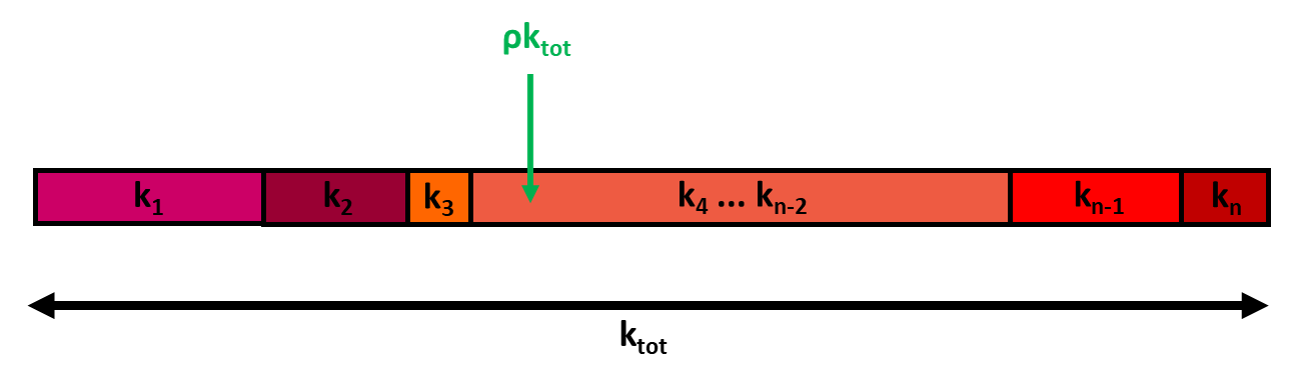}
\caption{Schematic respresentation of a kMC simulation. In the top, the different options for the system to leave a given state with the corresponding rates are depicted. Below, the kMC algorithm is graphically illustrated. By random number selection a value between 0 and k$_{tot}$ is generated, subsequently determining which process is chosen.}
\label{fig:KMC}
\end{figure}

Typically a system can leave a state by different pathways -- for a state $i$ there will exist several states $j$ the system can move to --  which are characterized by their individual rate constants $k_{ij}$ and an analogous escape probabilty $p_{ij}(t) = k_{ij} e^{-k_{ij} t}$,
with the rates of the single processes $k_{ij}$ summing up to the total rate $k_{tot} = \sum\limits_j k_{ij}$~\cite{Voter2007}. This agrees with the fact that an ensemble of independent Poisson processes can be reformulated as one Poisson process (see Eq.~(\ref{eq:escape}))~\cite{Fichthorn1991}. Now, instead of using the average time of escape of a process ($\tau = \int t\ p_{ij}(t) dt$) as kMC time step, a properly weighted stochastic escape time $\Delta t_{escape}$ has to be selected to guarantee a correct time evolution of the system. This can be achieved by reverting Eq.~(\ref{eq:survival}) and replacing $p_s(t)$ by a random number $\rho \in (0,1]$~\cite{Voter2007,Reuter2011,Andersen2019}:
\begin{equation}
\Delta t_{escape} = -\frac{ln(\rho)}{k_{tot}}
\end{equation}

To run a kMC simulation, one has to determine all $N$ processes that are possible for a given configuration of the system with the corresponding rates. Subsequently these rates are summed up to yield the overall rate $k_{tot}$.
Next, to select the process that will be executed, we plot the total escape rate as a bar of length $k_{tot}$, consisting of the bars representing the respective single rates (see Fig.~\ref{fig:KMC}). By multiplying $k_{tot}$ with a random number $\rho_1$, lying in the range (0,1], we end up in a certain bar, which then corresponds to the process that will be selected:
\begin{equation}
 \sum_{i=1}^{q} k_p \ \leq \ \rho_1 k_{tot} \ < \   \sum_{i=1}^{q-1} k_p
\end{equation}

 \begin{figure}[t]
  \center\includegraphics[angle=0,width=0.9\columnwidth]{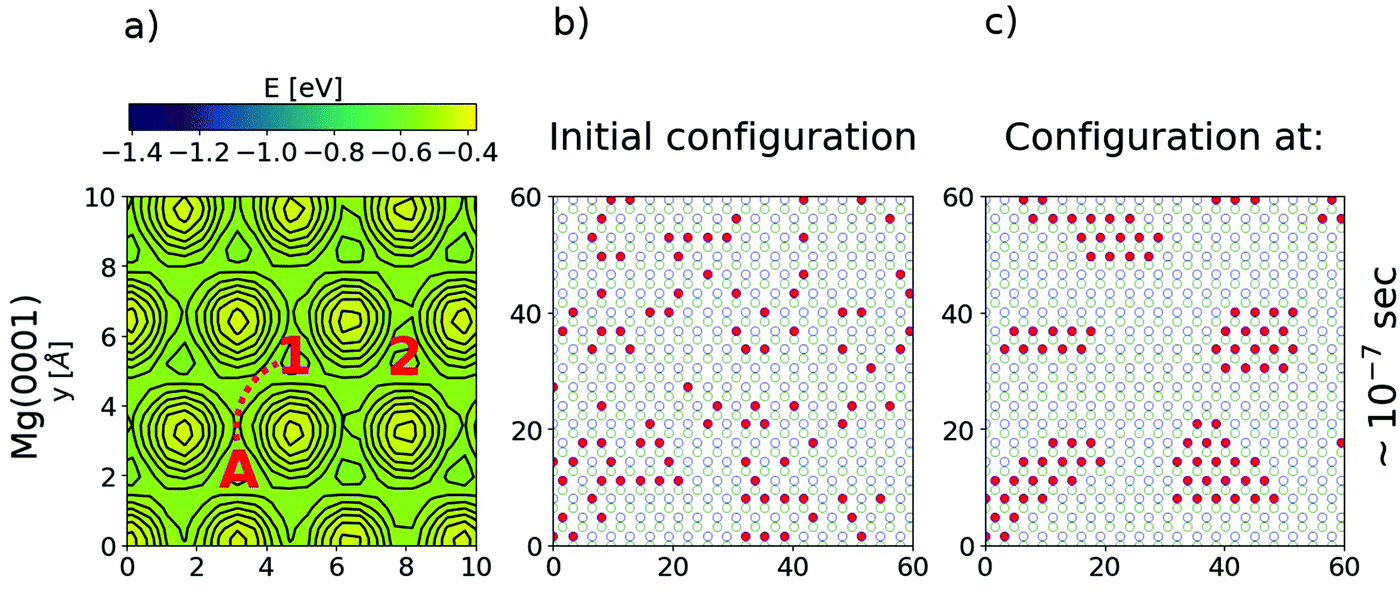}\vspace{-0.3cm}
    \center\includegraphics[angle=0,width=0.9\columnwidth]{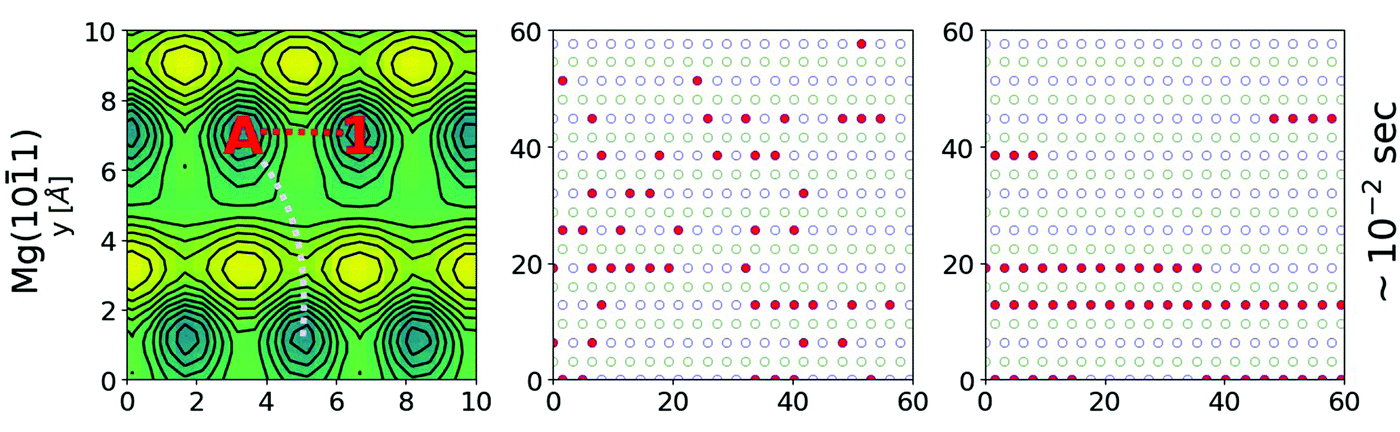}
 \caption{a) Potential energy landscape of two different Mg surfaces. Non-negligible interaction energies exist between adatom A and an additional adatom marked with a number. Red (pale pink) dotted lines represent hopping across lower (higher) energy barriers.  b) and c) show the initial random arrangement and a quasi--relaxed configuration that is obtained during a KMC simulation for Mg (0001) and the Mg (10$\overline{1}$1) surface after the denoted simulation time. Reproduced from Lautar et al.~\cite{Lautar2019}. With permission of the Royal Society of Chemistry}
 \label{fig:Mg-kMC}
\end{figure}

Finally, the simulation time is updated, \textit{i.e.}, $t = t + \Delta t$. For this purpose, the escape time $\Delta t = -\ln(\rho_2)/k_{tot}$ is obtained by choosing a second random number $\rho_2$. As discussed above, the choice of this time interval ensures a proper stochastic weighting of the time steps. Then the whole procedure restarts. This rejection free algorithm is often referred to as N--fold way and was originally designed to speed up Monte Carlo simulations, while it was only later used for kMC~\cite{Bortz1975,Kratzer2009}. With such a kMC approach, an efficient computational tool for the investigation of system dynamics is available. In particular, complex dynamics can be understood with respect to the underlying atomistic processes by applying kMC simulations~\cite{Mahlberg2021}. This is exemplified for a study on the morphology evolution of different facets of Mg surfaces by Lautar and co--workers~\cite{Lautar2019}. While Li--metal anodes are prone to shortcircuiting due to dendrite growth, this is typically not observed in Mg--batteries~\cite{Jaeckle2018}. Therefore, studies on the factors that influence the metal deposition are of great interest. This study points out that it is not sufficient to only investigate the most stable surface of a metal anode, since this surface typically does not account for the largest fraction of the overall surface area of a crystallite. Moreover, their nucleation theory based kMC studies show that different surfaces can indeed exhibit distinctly different growth modes. In Fig.~\ref{fig:Mg-kMC} the evolution on the most stable surface and the surface with the largest area fraction in the Wulff construction are depicted, which correspond to the Mg (0001) and Mg (10$\overline{1}$1) surface, respectively. The differences in their growth behaviour is nicely visible, with the Mg (0001) surface showing island growth, whereas the formation of lines is observed in case of the Mg (10$\overline{1}$1) surface.

\section{Conclusion}
In the first part of this review, we have given an concise overview on the basic concepts of a functioning battery, including the underlying electrochemical processes in cathode, anode and electrolyte.
The second part on the other hand has focused on the most prominent computational approaches that are available to model battery components on an atomistic scale. Here, a large part of the review was spent on DFT, a method which has become the work horse in many areas of material science. As a significant part of todays battery research is concerned with the search for improved materials, it is evident that DFT is an important tool for battery research. Combining high accuracy with efficient computation make DFT well--suited for large scale screening studies as well as for in depth investigations of particular systems, such that the importance of DFT is likely to even further increase. On the other hand, for thermodynamic information DFT methods (including AIMD) are often still not capable to treat the relevant length and time scales. For such problems, the more coarse--grained methods like cluster expansion based Monte Carlo or kinetic Monte Carlo approaches have been discussed. While these methods are built on different grounds, they are usually also based on input data from DFT. Finally, classical molecular dynamics simulations based on machine learning potentials have gained significance in recent years. Such potentials, again based on DFT data, will allow to study many problems that currently have not yet been addressed in sufficient detail. For instance, fundamental questions such as composition and formation of the famous solid electrolyte interface (SEI) may be addressed by such an approach. Consequently, an drastic increase of machine learning based molecular dynamics studies has to be expected, as a combination of DFT and machine learning may be used to study a large variety of otherwise intractable problems.

\section*{Acknowledgments}
This work has been supported by the Deutsche Forschungsgemeinschaft (DFG, German Research Foundation) under Germany´s Excellence Strategy – EXC 2154 – Project number 390874152 (POLiS Cluster of Excellence) and by the Dr. Barbara Mez-Starck Foundation. It contributes to the research performed at CELEST (Center for Electrochemical Energy Storage Ulm-Karlsruhe)

\bibliography{Review-all}{}

\end{document}